\documentclass[pdflatex,sn-mathphys-num]{sn-jnl}

\usepackage{graphicx}%
\graphicspath{{figures/}}%
\usepackage{multirow}%
\usepackage{amsmath,amssymb,amsfonts,amsthm}%
\usepackage{mathrsfs}%
\usepackage[title]{appendix}%
\usepackage{xcolor}%
\usepackage{textcomp}%
\usepackage{manyfoot}%
\usepackage{booktabs}%
\usepackage{algorithm}%
\usepackage{algorithmicx}%
\usepackage{algpseudocode}%
\usepackage{listings}%
\usepackage{tabularx}
\usepackage[dvipsnames]{xcolor}

\usepackage{lineno}
\usepackage{hyperref}

\usepackage{tikz}
\usetikzlibrary{shapes.geometric, arrows, arrows.meta, decorations.text}

\tikzstyle{startstop} = [rectangle, rounded corners,
minimum width=3cm,
minimum height=1cm,
text centered,
draw=black,
fill=VioletRed!50,
line width=1px]

\tikzstyle{process} = [rectangle,
minimum width=3cm,
minimum height=1cm,
align=center,
draw=black,
fill=Goldenrod!50,
line width=1px]

\tikzstyle{decision} = [diamond, 
aspect=2.5,
text centered,
draw=black,
fill=teal!40,
line width=1px,
inner sep=1.5pt]

\tikzstyle{arrow} = [line width=1px, -{Latex[length=7px, width=7px]}]

\newtheorem{remark}{Remark}%
\newtheorem{formulation}{Formulation}

\let\code\texttt

\newcommand{\mx}{n_{\mathcal{X}}}
\newcommand{\my}{n_{\mathcal{Y}}}
\newcommand{\na}{n_{\alpha}}
\newcommand{\nb}{n_{\beta}}
\newcommand{\nl}{n_{\ell}}

\newcommand{\ufem}{u_h}

\newcommand{\Omegax}{\Omega_{\mathcal{X}}}
\newcommand{\Omegay}{\Omega_{\mathcal{Y}}}
\newcommand{\omegax}{\omega_{\mathcal{X}}}
\newcommand{\omegay}{\omega_{\mathcal{Y}}}
\newcommand{\Gammad}{\Gamma_{\hspace{-.02in} d}}
\newcommand{\Gamman}{\Gamma_{\hspace{-.02in} n}}
\newcommand{\Gammal}{\Gamma_{\hspace{-.02in} \alpha}}
\newcommand{\Gammar}{\Gamma_{\hspace{-.02in} \beta}}

\newcommand{\ind}[1]{\boldsymbol{1}_{#1}}

\newcommand{\Fr}{F_{_{\hspace{-.02in} R}}}
\newcommand{\aerro}{\hat{\epsilon}_{\omega}}
\newcommand{\aerrg}{\hat{\epsilon}_{\gamma}}
\newcommand{\perrg}{\epsilon_{\gamma}}
\newcommand{\cfom}{c_{\text{\itshape FOM}}}
\newcommand{\cfac}{c_{\text{\itshape fac}}}
\newcommand{\csub}{c_{\text{\itshape sub}}}
\newcommand{\crom}{c_{\text{\itshape ROM}}}
\newcommand{\coff}{c_{\text{\itshape off}}}
\newcommand{\con}{c_{\text{\itshape on}}}
\newcommand{\tfac}{t_{\text{\itshape fac}}}
\newcommand{\tsub}{t_{\text{\itshape sub}}}
\newcommand{\toff}{t_{\text{\itshape off}}}
\newcommand{\ton}{t_{\text{\itshape on}}}

\begin{document}

\title[Accelerated estimation of quantities of interest via adjoint-based model reduction]{Accelerated estimation of quantities of interest\par via adjoint-based model reduction}


\author*[1]{\fnm{Cl\'{e}ment} \sur{Vella}}\email{cvella@suqaba.com}

\author[2]{\fnm{Serge} \sur{Prudhomme}}\email{serge.prudhomme@polymtl.ca}

\affil[1]{\orgname{Suqaba}, \orgaddress{\street{47 Avenue Victor Hugo}, \city{Vanves}, \postcode{92170}, \country{France}}}

\affil[2]{\orgdiv{Department of Mathematics and Industrial Engineering},\par
\orgname{Polytechnique Montr\'{e}al},\par
\orgaddress{\street{C.P.~6079, succ.~Centre-ville}, \city{Montr\'{e}al}, \postcode{H3C 3A7}, \state{Qu\'ebec}, \country{Canada}}}

\abstract{We introduce an adjoint-based reduced-order model framework for fast and accurate estimation of quantities of interest for \emph{many-query} linear problems. The method builds a reduced-order model with respect to the adjoint problem, thus bypassing the solution of the primal problem and drastically reducing computational cost. It creates a surrogate model that is independent of the loading configurations. It enables fast evaluation across multiple load cases and the generation of virtual charts to support decision-making. Numerical experiments on the Poisson equation and a plane-stress elasticity problem demonstrate that the adjoint reduced-order model converges rapidly, outperforms its primal counterpart, and provides reliable estimates of the quantities of interest. Importantly, it is often more practical to parameterize a kernel function than an entire set of external loads, making the method generic and particularly suited for early-stage prototyping and design optimization.}



\maketitle

\section{Introduction}
\label{sec:intro}

Problems in design, control, optimization, or uncertainty quantification have substantially increased computational costs~\cite{benner2015survey}. This is driven by the demand for more predictive simulations, which requires accounting for variability in boundary conditions, material properties, and geometry. Reduced-Order Modeling (ROM) provides a framework to address this challenge, offering compact representations of high-fidelity models to enable fast evaluations of model outputs. Among the most widely used approaches is the Proper Orthogonal Decomposition (POD), which constructs low-dimensional bases from snapshots of high-fidelity simulations. It is particularly effective when representative training data are available~\cite{chatterjee2000introduction, kunisch2001galerkin, gunzburger2007reduced, lu2019review}. The Reduced Basis (RB) method extends the concept by incorporating error estimation and adaptive sampling, enabling certified reduction in parameterized settings~\cite{maday2002reduced, haasdonk2008reduced, quarteroni2015reduced, haasdonk2017reduced}. The Proper Generalized Decomposition (PGD) follows a different paradigm: it constructs an \emph{a priori} separated representation directly from the variational formulation, allowing efficient exploration of high-dimensional parameter spaces without relying on snapshots~\cite{chinesta2013proper, ladeveze2014pgd, chinesta2013pgd}.

In many applications, one is often interested in specific quantities of interest (QoIs) of the solution rather than in the full solution field. This has motivated the development of goal-oriented error estimation techniques~\cite{gartland1984computable, becker1996feed, paraschivoiu1997posteriori, prudhomme1999goal, oden2001goal,almeida2010dual}, which rigorously control errors with respect to targeted outputs. Coupled with ROM techniques, this perspective enables surrogate models that remain computationally efficient while being tailored to accurately predict the desired QoIs. Several goal-oriented extensions of reduced-order methods have been proposed. A goal-oriented variant of POD was presented in~\cite{carlberg2011low}, where additional snapshots based on parametric derivatives (so-called sensitivity factors) were incorporated. This approach produces reduced-order solutions that capture parameter variations more accurately than the classical POD. Within the RB framework, goal-oriented strategies have been extensively investigated for a range of problems, including steady linear PDEs~\cite{prud2002reliable}, unsteady linear parabolic PDEs~\cite{prud2002reduced, rovas2006reduced}, and nonlinear fluid dynamics~\cite{carlberg2017galerkin}. \emph{A posteriori} error bounds are central to these approaches, and are specifically designed to estimate the error in QoIs. Dual-Weighted Residual (DWR) techniques have played a central role in extending goal-oriented error estimation to model reduction. Within the RB framework, DWR-based estimators are frequently embedded in greedy-sampling strategies to guide the selection of full-order solves and thereby enrich the reduced basis~\cite{nguyen2007posteriori, rozza2008reduced, fischer2024more}. Beyond sampling, DWR has been leveraged for adaptive refinement of reduced models: $h$-adaptation of the basis was proposed in~\cite{carlberg2015adaptive, etter2020online}, while joint adaptation of the basis and hyper-reduction sample points was demonstrated in~\cite{collins2019}. In the context of hyper-reduction, reduced quadrature rules have been developed from DWR formulations~\cite{yano2020goal, sleeman2022goal}. In~\cite{meyer2003efficient}, goal-oriented estimates were employed to eliminate POD basis vectors with negligible contribution to the accuracy of the quantity of interest. Goal-oriented strategies have also been developed within the PGD framework, primarily with the aim of verification and certification~\cite{alfaro2015error, de2013basis, ammar2010error, chamoin2017posteriori, ladeveze2011verification}. A distinctive approach was introduced in~\cite{kergrene2019goal}, where a reduced adjoint solution is computed first and incorporated directly into the solution of the primal PGD problem as a constraint on the error in the quantity of interest, rather than subsequently solving the adjoint problem for error estimation. It was shown that this constrained strategy improved the accuracy of QoI predictions compared to standard PGD formulations.

This work focuses on problems involving many loading configurations on a given geometry. Modern engineering workflows frequently require \emph{many-query} analyses, where structures are evaluated under a large number of load cases. Examples include generative design and rapid prototyping, which demand repeated assessment of candidate designs under varying loads, and probabilistic risk assessment, which relies on extensive simulations of structural responses to uncertain inputs. Although not explicitly goal-oriented, a PGD framework has been developed for nonlinear dynamic systems subjected to multiple non-parameterized loading conditions. The method employs a ``parent simulation'' strategy, where similar load cases are identified so that reduced bases and initializations can be reused, thereby accelerating simulations in many-query settings~\cite{daby2025model}. On another note, various Krylov subspace methods have been developed at the linear solver level to exploit common structures across repeated solves. Block-Krylov solvers~\cite{o1980block} and recycling/deflation schemes~\cite{farhat1994extending, saad2000deflated, erhel2000augmented} reuse Krylov subspaces or previously converged modes when solving successive right-hand sides. These strategies have also been incorporated into multigrid and domain decomposition solvers to handle multiple right-hand sides simultaneously~\cite{jolivet2016block}. Despite these advances, repeated solves of large-scale systems remain costly, further motivating the use of reduced-order models.

In this work, a surrogate-based approach for linear boundary-value problems is developed by performing model reduction only on the associated adjoint problem. 
This enables 1)~the construction of a ROM surrogate for the adjoint solution, and 2)~evaluation of the QoI for any load configuration via the surrogate adjoint solution. In effect, the primal problem is bypassed: all computational effort is concentrated in the adjoint-based surrogate, and any query reduces to inexpensive post-processing. It will be demonstrated that this approach preserves accuracy while dramatically reducing online computation compared with solving each load case using a full-order solver. Furthermore, results will show that the adjoint-based surrogate not only achieves computational savings but also provides improved accuracy, compared to a primal-based surrogate. We demonstrate the developed method on two test cases: 1)~Poisson equation with non-parameterized source terms, serving as a proof-of-concept to verify the accuracy of the surrogate; 2)~a two-dimensional plane-stress elasticity problem, representative of an industrial design task, where the structure is subjected to many load configurations. In this case, we show that the adjoint-PGD surrogate can compute QoI estimates with high fidelity while drastically reducing the cost per evaluation.

The remainder of the paper is organized as follows. Section~\ref{sec:sec2} introduces the model problems under consideration. Section~\ref{sec:pgd} presents the Proper Generalized Decomposition (PGD), including the adjoint-based surrogate modeling and multi-query framework. Section~\ref{sec:num_ex} reports numerical examples to assess accuracy and computational performance, and discusses possible extensions. Finally, Section~\ref{sec:conclusion} provides concluding remarks.

\section{Model problems}
\label{sec:sec2}

\subsection{Poisson problem}
\label{subsec:poisson}

Let~$\Omega \in \mathbb R^2$ be an open bounded domain with Lipschitz boundary~$\partial \Omega$. The solution field~$u : \bar{\Omega} \rightarrow \mathbb{R}$, subjected to homogeneous Dirichlet boundary conditions, is governed by the following set of equations:
\begin{equation}
\label{eq:poisson_eq}
\begin{aligned}
- \Delta u = f, &\quad \text{in}\ \Omega, \\
u = 0, &\quad \text{on}\ \partial \Omega.
\end{aligned}
\end{equation}
The function~$f : \Omega \rightarrow \mathbb{R}$ is supposed to be sufficiently regular to yield a well-posed problem. We also introduce the Hilbert spaces: 
\[
\begin{aligned}
&H^1(\Omega)=\{v \in L^2(\Omega)\ \vert\ \nabla v \in (L^2(\Omega))^2\}, \\ 
&H_0^1(\Omega)=\{v \in H^1(\Omega)\ \vert \ v = 0 \ \text{on}\ \partial \Omega \}, \\
&V = H_0^1(\Omega).
\end{aligned}
\]

\begin{formulation}[Weak formulation of the Poisson problem]
\label{frm:poisson}
The weak form of the Poisson problem reads:
\[
\label{eq:weak_steadyheat}
\text{Find}\ u \in V\ \text{such that}\ 
\int_{\Omega}{ \nabla u \cdot \nabla v \, dx } 
= \int_{\Omega}{ f v \, dx }, \quad \forall v \in V, \]
which can be recast as:
\begin{equation}
\text{Find $u \in V$, such that} \ B(u, v) = F(v), \quad \forall v \in V,
\end{equation}
where the bilinear form $B$ and linear form $F$ are defined as:
\[
\begin{aligned}
&B(u, v) = \int_{\Omega}{ \nabla u \cdot \nabla v \, dx }, \\
&F(v) = \int_{\Omega}{ f v \, dx } .
\end{aligned}
\]
\end{formulation}

\noindent 
We suppose in this work that one is interested in the solution $u$ only at some points $\mu$ within a subdomain $\omega \subset \Omega$ rather than on the whole computational domain $\Omega$. In other words, we would consider a family of quantities of interest (QoI) defined as, for $u$ sufficiently smooth:
\[
Q_{\mu}(u) = u(\mu) = \int_{\Omega}{\delta(x-\mu) u(x)\, dx},
\quad \mu \in \omega \subset \Omega,
\]
where $\delta$ is the Dirac delta function. 
However, it is well known that the linear functionals $Q_{\mu}(v)$ given above are not necessarily bounded for $v \in V$ and would lead to an ill-posed adjoint problem. In practice, we will thus replace the Dirac function by a kernel function $k_\epsilon$, where $\epsilon$ indicates that its support is small compared to the size of $\Omega$, so that the quantities of interest:
\[
Q_{\mu}(u) = u(\mu) = \int_{\Omega}{k_\epsilon(x-\mu) u(x)\, dx},
\quad \mu \in \omega \subset \Omega,
\]
define bounded linear functionals on $V$. The idea will be to choose the kernel function in such a way that the quantities of interest can be viewed as averages of the solution $u$ over small subdomains. 

\begin{formulation}[Weak formulation of the adjoint problem to Formulation~\ref{frm:poisson}]
\label{frm:adj_poisson}
The weak form of the adjoint problem associated with the quantity of interest $Q_{\mu}(u)$ reads:
\[
\text{Find}\ z_{\mu} \in V\ \text{such that}\ 
B(v, z_{\mu}) = Q_{\mu}(v), \quad \forall v \in V,
\]
that is:
\begin{equation}
\label{eq:weak_adj_poisson}
\text{Find}\ z_{\mu} \in V\ \text{such that}\ \int_{\Omega}{ \nabla v \cdot \nabla z_{\mu} \, dx } = \int_{\Omega}{ k_\epsilon(x-\mu) v(x)\, dx}, \quad \forall v \in V. 
\end{equation}
\end{formulation}

\noindent 
Whenever the adjoint solution $z_{\mu}$ is sufficiently smooth, one can integrate by parts the integral on the left-hand side of the equation to obtain the strong formulation of the adjoint problem:
\[
\begin{aligned}
- \Delta z_{\mu} = k_\epsilon(x-\mu), & \quad \forall x \in \Omega,\\
z_{\mu}(x) = 0, & \quad \forall x \in \partial \Omega .
\end{aligned}
\]
The adjoint solution, parameterized by $\mu$, is therefore governed by the Poisson equation and subjected to homogeneous Dirichlet conditions and loading term $k_\epsilon(x-\mu)$.

\subsection{Plane-stress elasticity}
\label{subsec:pse}

The second model problem we consider deals with plane-stress elasticity. Let~$\Omega \subset \mathbb{R}^{2}$ be an open bounded subset with Lipschitz boundary~$\partial\Omega = \overline{\Gammad \cup \Gamman}$, where~$\Gammad$ and~$\Gamman$ are disjoint portions of the boundary associated with Dirichlet and Neumann conditions, respectively. The displacement field~$u : \bar{\Omega} \rightarrow \mathbb{R}^{2}$ satisfies the following set of equations:
\[
\begin{aligned}
- \nabla \cdot \sigma(u) = f, 
& \quad \text{in}\ \Omega, \\
\sigma(u) = \mathbf{C} : \varepsilon(u), 
& \quad \text{in}\ \Omega, \\
\varepsilon(u) = \frac{1}{2} \left(\nabla u + (\nabla u)^{\top} \right), 
& \quad \text{in}\ \Omega, \\
u = 0, 
& \quad \text{on}\ \Gammad, \\
n \cdot \sigma(u) = g, 
& \quad \text{on}\ \Gamman,
\end{aligned}
\]
where~$f : \Omega \rightarrow \mathbb{R}^{2}$ represents the body force density and $g : \Gamman \rightarrow \mathbb{R}^{2}$ defines the prescribed traction density on $\Gamman$. The vector~$n$ denotes the unit outward normal to the boundary. The stress tensor~$\sigma(u)$ is related to the strain tensor~$\varepsilon(u)$ according to Hooke's law, which, in the case of isotropic materials, reads:
\[
\sigma(u) = 2 \mu \, \varepsilon(u) + \lambda \, tr \big(\varepsilon(u) \big) \, I_{2},
\]
where~$\lambda$ and~$\mu$ are the Lam\'{e} parameters and $I_{2} \in \mathbb{R}^{2 \times 2}$ is the identity tensor. Under plane stress conditions and using the Voigt form, the Hooke's law $\sigma(u) = \mathbf{C} : \varepsilon(u)$ in matrix notation reads:
\begin{equation}
\label{eq:HookesLaw}
\begin{bmatrix}
\sigma_{11} \\ \sigma_{22} \\ \sigma_{12}
\end{bmatrix} %
= %
\begin{bmatrix}
2 \mu + \lambda &         \lambda &   0 \\
\lambda         & 2 \mu + \lambda &   0 \\
0               & 0               & \mu
\end{bmatrix} %
\begin{bmatrix}
\varepsilon_{11} \\ \varepsilon_{22} \\ 2 \varepsilon_{12}
\end{bmatrix} .
\end{equation}

\begin{formulation}[Weak formulation of the plane-stress elasticity problem]
\label{frm:ps_elas}
\begin{equation}
\label{eq:weak_elasticity}
\text{Find}\ u \in U\ \text{such that}\ \int_{\Omega}{ \sigma(u) : \varepsilon(v) \, dx } = \int_{\Omega}{ f \cdot v \, dx } + \int_{\Gamman}{ g \cdot v \, ds }, \quad \forall v \in U, 
\end{equation}
with:
\[
U = \{ v \in [H^{1}(\Omega)]^{2}\ \vert\ v = 0\ \text{on}\ \Gammad \}.
\]
Equation~\eqref{eq:weak_elasticity} can be recast as:
\[
\text{Find}\ u \in U \ \text{such that}\ A(u, v) = L(v), \quad \forall v \in U,
\]
where the bilinear form and linear form are given by:
\[
\begin{aligned}
&A(u, v) = \int_{\Omega}{\sigma(u) : \varepsilon(v) \, dx}, \\
&L(v) = \int_{\Omega}{f \cdot v \, dx + \int_{\Gamman} g \cdot v \, ds} .
\end{aligned}
\]
\end{formulation}

\noindent
We suppose here that we are interested in the evaluation of the normal displacement $u\cdot n$ at some points $\mu$ located on a sub-portion $\gamma \subset \Gamman$ of the boundary. We will consider the following QoI:
\[
J_{\mu}(u) = \int_{\Gamman}{ k_{\epsilon}(x-\mu) \, n(x) \cdot u(x)\, ds }, \quad \mu \in \gamma,
\]
where $k_{\epsilon}$ is a kernel function defined on $\Gamman$.
The kernel function will be chosen such that $J_{\mu}(u)$ is a local average of the normal displacement to the boundary~$\gamma$.

\begin{formulation}[Weak formulation of the adjoint problem to Formulation~\ref{frm:ps_elas}]
\label{frm:adj_ps_elas}
\[
\text{Find}\ z_{\mu} \in U\ \text{such that}\ 
A(v, z_{\mu}) = J_{\mu}(v), \quad \forall v \in U,
\]
which also reads:
\begin{equation}
\label{eq:weak_adj_ps_elas}
\text{Find}\ z_{\mu} \in U\ \text{such that}\ \int_{\Omega}{ \sigma(v) : \varepsilon(z_{\mu}) \, dx } = \int_{\Gamman}{ k_{\epsilon}(x-\mu) \, n(x) \cdot v(x)  \, ds }, \quad \forall v \in U. 
\end{equation}
\end{formulation}

\noindent 
Assuming that the adjoint is sufficiently smooth to be able to apply the divergence theorem and using the symmetry of the elasticity tensor $\mathbf C$, one obtains the strong formulation of the adjoint problem as:
\[
\begin{aligned}
- \nabla \cdot \sigma(z_{\mu}) = 0,
& \quad \text{in}\ \Omega, \\
\sigma(z_{\mu}) = \mathbf{C} : \varepsilon(z_{\mu}), 
& \quad \text{in}\ \Omega, \\
\varepsilon(z_{\mu}) = \frac{1}{2} \left(\nabla z_{\mu} + (\nabla z_{\mu})^{\top} \right), 
& \quad \text{in}\ \Omega, \\
z_{\mu}(x) = 0, 
& \quad \forall x \in \Gammad, \\
\sigma(z_{\mu})(x) \cdot n(x) = k_{\epsilon}(x-\mu) \, n(x), 
& \quad \forall x \in \Gamman.
\end{aligned}
\]
The adjoint solution $z_\mu$, parameterized by~$\mu$, is governed in the case of the quantity of interest $J_\mu(u)$ by an equilibrium equation with a homogeneous right-hand side  and subjected to homogeneous Dirichlet boundary conditions on $\Gammad$ and Neumann conditions on $\Gamman$.

\section{Proper Generalized Decomposition}
\label{sec:pgd}

\subsection{Adjoint-based surrogate modeling}

In many applications, such as generative design or fast prototyping, one must evaluate the response of a given system for a large number of loading configurations. Since optimization and analysis often focus on a specific quantity of interest (QoI), it seems natural to construct surrogate models associated with the adjoint problem, which captures the sensitivity of the QoI with respect to variations in the loading configurations, rather than repeatedly solving the primal problem for each load and then computing the QoI.

\begin{figure}[tb]
\centering
\includegraphics[width=.9\textwidth]{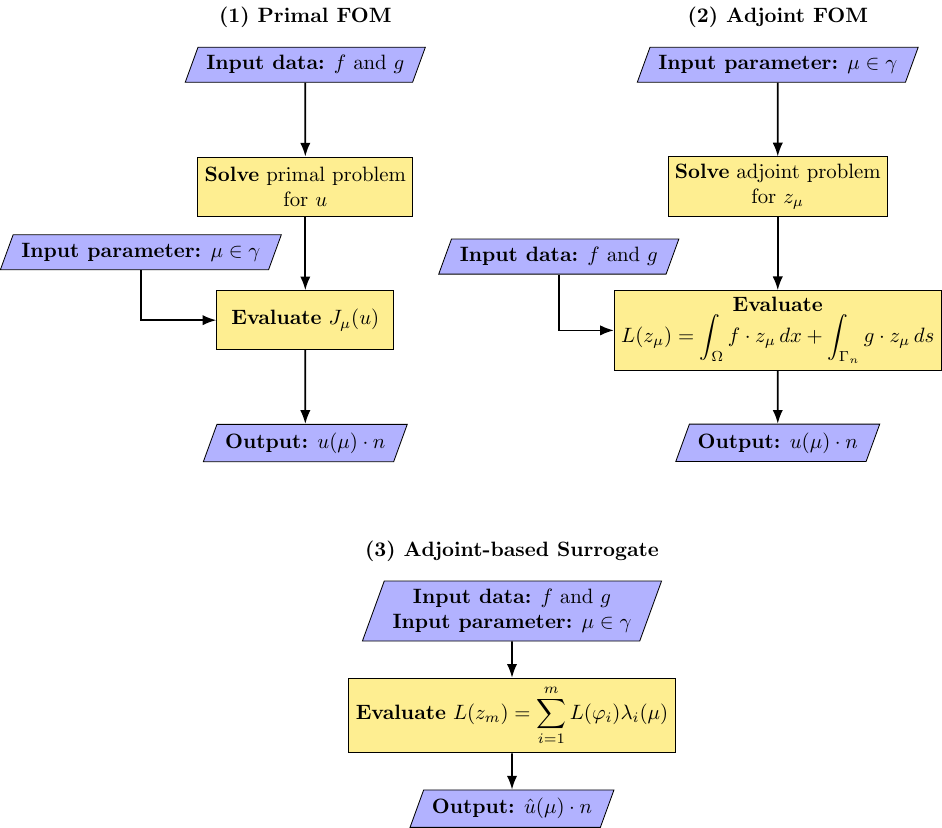}
\caption{Flowcharts of the three approaches to compute QoI estimates for the plane-stress elasticity problem, namely (1) the Primal Full-Order Model; (2) the Adjoint Full-Order Model; and (3) the Adjoint-based Surrogate.}
\label{fig:primal_dual_surrogate}
\end{figure}

A key observation is that, in the current framework, the adjoint problem is independent of the external loads. This makes the method generic, regardless of whether the external loads are parameterized. As a trade-off, the complexity is shifted to the evaluation of the adjoint solution across multiple values of the parameter~$\mu$. In the presented framework, which involves linear problems and linear QoI, the evaluation of the QoI estimates will be performed using the fundamental relations:
\[
Q_{\mu}(u) = B(u,z_{\mu}) = F(z_{\mu}) ,
\]
for the Poisson equation, or similarly,
\[
J_{\mu}(u) = A(u,z_{\mu}) = L(z_{\mu}) ,
\]
for the plane-stress elasticity problem.
The efficient construction of surrogate models for the adjoint problems would enable rapid assessment of the response of the system for a large number of external loads. The idea is thus to bypass the computation of the solution to the primal problem to evaluate $Q_{\mu}$ and $J_{\mu}$. Instead, estimates $\hat{z}_{\mu}$ of the adjoint solution $z_{\mu}$ can efficiently be obtained via surrogate modeling to evaluate $F(\hat{z}_{\mu})$ and $L(\hat{z}_{\mu})$ as estimates of the QoI (see Figure~\ref{fig:primal_dual_surrogate}).

We shall use in this work the Proper Generalized Decomposition (PGD) to construct the reduced-order model associated with the adjoint problem. The technique is described for the Poisson problem and the plane-stress elasticity equations in the following sections.

\subsection{Poisson equation}

\begin{figure}[tb]
\centering
\includegraphics[width=.35\textwidth]{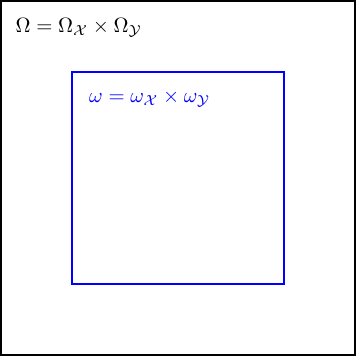}
\caption{Schematic of the 2D computational domain $\Omega$ and subdomain $\omega \subset \Omega$, in which the solution to the Poisson equation is of interest.}
\label{fig:poisson_omega}
\end{figure}

We assume that the domain~$\Omega$ is the tensor product of two one-dimensional intervals~$\Omegax$ and~$\Omegay$, i.e., $\Omega = \Omegax \times \Omegay$. Similarly, the subset~$\omega$ is decomposed as $\omega = \omegax \times \omegay$, as shown in Figure~\ref{fig:poisson_omega}. Moreover, we assume that the kernel function $k_\epsilon$ is chosen as
\[
k_{\epsilon}(x-\mu_x, y-\mu_y) = \frac{1}{2 \pi \epsilon^{2}} \, \exp \left( -\frac{(x - \mu_{x})^{2}}{2 \epsilon^{2}} \right) \, \exp \left( -\frac{(y - \mu_{y})^{2}}{2 \epsilon^{2}} \right),
\]
with~$(\mu_{x}, \mu_{y}) \in \omegax \times \omegay$ and~$\epsilon \in \mathbb{R}_{+}^{\ast}$.
Although the Gaussian kernels are not compactly supported, we shall assume that they can be truncated as needed resulting in negligible errors in the evaluation of the quantities of interest. Moreover, we observe that they are not readily separable with respect to~$x$ and $\mu_{x}$, nor to $y$ and $\mu_{y}$, but we shall assume the existence of separated representations such that:
\begin{equation}
\label{eq:poisson_gk}
k_{\epsilon}(x - \mu_x, y - \mu_y) 
\simeq \bar{k}_\epsilon (x-\mu_{x},y-\mu_{y}) =
\bigg( \sum_{\kappa = 1}^{\mx}{ g_{\kappa}(x) \, h_{\kappa}(\mu_{x}) } \bigg) 
\bigg( \sum_{j = 1}^{\my}{k_{j}(y) \, \ell_{j}(\mu_{y})} \bigg).
\end{equation}
In practice, we will perform a Singular Value Decomposition (SVD) on the discrete kernel, where~$g_{\kappa}$ and~$k_{j}$ are the spatial basis functions and~$h_{\kappa}$ and~$\ell_{j}$ denote the parametric basis functions. The number of modes $\mx$ and $\my$ with respect to the two directions will be chosen so as to minimize the decomposition errors. The reader may refer to~\cite{zlotnik2015} for the study of approximation errors due to non-separable input data in the PGD framework for the Poisson equation.

\begin{formulation}[Weighted residual formulation of the parameterized adjoint problem associated with the Poisson equation]
\label{frm:param_wr_poisson}
\[
\text{Find}\ z_{\mu} \in V\times L^{2}(\omega)\ \text{such that}\ \int_{\omega}{ B(v, z_{\mu}) \, d\mu} = \int_{\omega}{ Q_{\mu}(v) \, d\mu}, \quad \forall v \in V\times L^{2}(\omega) . 
\]
\end{formulation}

\noindent 
The PGD method consists then in searching for an approximation $z_{m}$ of the adjoint solution $z_\mu$ to the above problem in the form:
\[
z_{m}(x, y, \mu_{x}, \mu_{y}) = \sum_{i = 1}^{m}{ 
\varphi_{i}(x) \,
\psi_{i}(y) \,
\lambda_{i}(\mu_{x}) \,
\eta_{i}(\mu_{y}) }.
\]
The separated representation is built progressively by adding a single rank-one term at each enrichment step, following a greedy update algorithm. More precisely, given the approximation~$z_{m-1}$, we compute the next new modes at the enrichment step~$m$ as:
\[
z_{m} = \left( \sum_{i = 1}^{m - 1}{ \varphi_{i} \psi_{i} \lambda_{i} \eta_{i} } \right) + \varphi \psi \lambda \eta = z_{m - 1} + \varphi \psi \lambda \eta ,
\]
where we have dropped the subscript $m$ in $\varphi_m$, $\psi_m$, $\lambda_m$, and $\eta_m$ for the sake of clarity in the notation. Moreover, for the first enrichment step $m=1$, we choose the initial guess as $z_0 = 0$. We now rewrite Formulation~\ref{frm:param_wr_poisson} in terms of the PGD approximation using the fact that~$B(\cdot, \cdot)$ is bilinear.

\begin{formulation}[PGD formulation of the parameterized adjoint problem associated with the Poisson equation]
\label{frm:pgd_wr_poisson}
\[
\begin{aligned}
& \text{Find}\ (\varphi, \psi, \lambda, \eta) \in V_{\mathcal{X}} \times V_{\mathcal{Y}} \times L^{2}(\omegax) \times L^{2}(\omegay)\ \text{such that} \\
& \hspace{.5in} \int_{\omega}{ B(v, \varphi \psi \lambda \eta) \, d\mu} = \int_{\omega}{ Q_{\mu}(v) \, d\mu} - \int_{\omega}{ B(v, z_{m-1}) \, d\mu}, \quad \forall v \in V \times L^{2}(\omega),
\end{aligned}
\]
where the spaces $V_{\mathcal{X}}$ and $V_{\mathcal{Y}}$ are defined as:
\[
\begin{aligned}
& V_{\mathcal{X}} = \big\{ v \in H^1(\Omega_{\mathcal{X}})\ \vert\ v = 0\ \text{on}\ \partial \Omega_{\mathcal{X}} \big\}, \\
& V_{\mathcal{Y}} = \big\{ v \in H^1(\Omega_{\mathcal{Y}})\ \vert\ v = 0\ \text{on}\ \partial \Omega_{\mathcal{Y}} \big\}.
\end{aligned}
\]
\end{formulation}

\begin{figure}[tb]
\centering
\begin{tikzpicture}[node distance=1.55cm]

\node (start) [startstop] {Start fixed-point};

\node (init) [process, below of=start] {%
	$k \leftarrow 0$ \\ Initialize $\psi^{(0)}$, $\lambda^{(0)}$ and $\eta^{(0)}$};

\node (spgd) [process, below of=init, yshift=-.1in] {%
Given $\psi^{(k)}$, $\lambda^{(k)}$ and $\eta^{(k)}$, \\ solve Formulation~\ref{frm:poisson_pgd_x} for $\varphi^{(k+1)}$ \\
and Aitken acceleration on $\varphi^{(k+1)}$};

\node (apgd) [process, below of=spgd, yshift=-.1in] {%
Given $\varphi^{(k+1)}$, $\lambda^{(k)}$ and $\eta^{(k)}$, \\ solve Formulation~\ref{frm:poisson_pgd_y} for $\psi^{(k+1)}$};

\node (bpgd) [process, below of=apgd] {%
Given $\varphi^{(k+1)}$, $\psi^{(k+1)}$ and $\eta^{(k)}$, \\ solve Formulation~\ref{frm:poisson_pgd_mux} for $\lambda^{(k+1)}$};

\node (tpgd) [process, below of=bpgd] {%
Given $\varphi^{(k+1)}$, $\psi^{(k+1)}$ and $\lambda^{(k+1)}$, \\ solve Formulation~\ref{frm:poisson_pgd_muy} for $\eta^{(k+1)}$};

\node (conv) [decision, below of=tpgd, yshift=-.1in] {convergence?};

\node (kplus) [process, left of=bpgd, xshift=-1.25in] {$k \leftarrow k + 1$};

\node (stop) [startstop, below of=conv, yshift=-.1in] {Stop fixed-point};

\draw [arrow] (start) -- (init);
\draw [arrow] (init) -- (spgd);
\draw [arrow] (spgd) -- (apgd);
\draw [arrow] (apgd) -- (bpgd);
\draw [arrow] (bpgd) -- (tpgd);
\draw [arrow] (tpgd) -- (conv);
\draw [arrow] (conv.west) -| (kplus.south);
\draw [arrow] (kplus.north) |- (spgd.west);
\node (false) [left of=conv, xshift=-.1in, yshift=-.12in] {false};
\draw [arrow] (conv) -- node[anchor=south, xshift=.5cm, yshift=-.1cm] {true} (stop);

\end{tikzpicture}
\caption{Flowchart of the fixed-point algorithm for the PGD modes.}
\label{fig:flowchart_fixedpoint}
\end{figure}
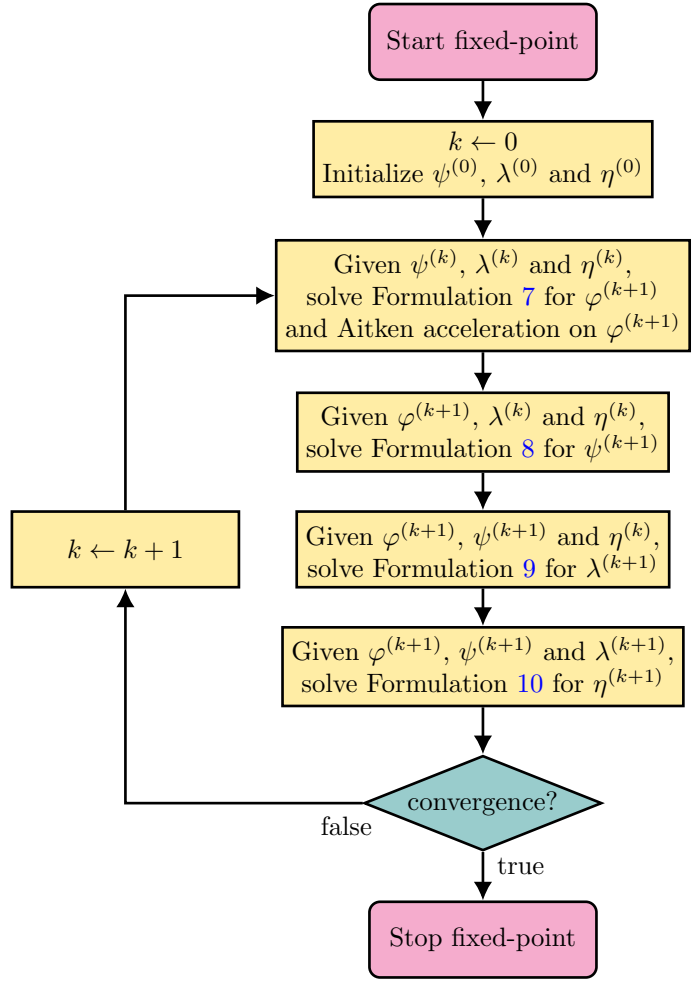

\noindent 
For each rank, Formulation~\ref{frm:pgd_wr_poisson} is nonlinear in the unknowns~$(\varphi, \psi, \lambda, \eta)$ as test functions in $ V\times L^{2}(\omega)$ will be chosen in the form:
\[
\begin{aligned}
v(x,y,\mu_x,\mu_y) 
&= \varphi^{\ast}(x) \, \psi(y) \, \lambda(\mu_{x}) \, \eta(\mu_{y})
+\varphi(x) \, \psi^{\ast}(y) \, \lambda(\mu_{x}) \, \eta(\mu_{y})\\
&+\varphi(x) \, \psi(y) \, \lambda^{\ast}(\mu_{x}) \, \eta(\mu_{y})
+\varphi(x) \, \psi(y) \, \lambda(\mu_{x}) \, \eta^{\ast}(\mu_{y}),
\end{aligned}
\]
where $\varphi^{\ast} \in V_{\mathcal{X}}$, $\psi^{\ast} \in V_{\mathcal{Y}}$, $\lambda^{\ast} \in L^{2}(\omegax)$, and $\eta^{\ast} \in L^{2}(\omegay)$ are test functions in the respective one-dimensional spaces.
Therefore, the above formulation is usually solved by means of a fixed-point algorithm. In the end, the approach significantly reduces computational complexity by decoupling the problem into one-dimensional subproblems involving one-dimensional integrals only, see for instance the expression of $B(\cdot, \cdot)$ in Appendix~\ref{apx:eval_B}. The process of solving for the modes in the PGD approximation is referred to as the \emph{offline} or \emph{training phase}. The fixed-point iteration is illustrated in Figure~\ref{fig:flowchart_fixedpoint}. Note that the sequence $\left( \varphi^{(k)} \right)_{k \in \mathbb{N}}$ is accelerated by means of the Aitken's $\Delta^{2}$ process~\cite{aitken1932xii}. The reader is referred to~\cite{nouy2018multiscale,vella2024efficient,vella2025proper} for a detailed description of the implementation of the Aitken's acceleration method into the PGD fixed-point procedure. We now established each PGD subproblem below.

\begin{formulation}[PGD problem in parameter~$x$ for the Poisson equation]
\label{frm:poisson_pgd_x}
Assuming $\psi$, $\lambda$, and $\eta$ to be known and taking test functions in the form $v(x,y,\mu_x,\mu_y)  = \varphi^{\ast}(x) \psi(y) \lambda(\mu_{x}) \eta(\mu_{y})$, the weak formulation of the adjoint problem reads:
\[
\text{Find}\ \varphi \in V_{\mathcal{X}}\ \text{such that} \quad B_{\mathcal{X}}(\varphi, \varphi^{\ast}) 
= \mathcal R_{\mathcal{X}}(\varphi^{\ast}), 
\quad \forall \varphi^{\ast} \in V_{\mathcal{X}},
\]
where:
\[
\begin{aligned}
&B_{\mathcal{X}}(\varphi, \varphi^{\ast}) = \left( \int_{\omegax}{ \lambda^{2} \, d\mu_{x} } \right) \left( \int_{\omegay}{ \eta^{2} \, d\mu_{y} } \right) B(\varphi^{\ast} \psi, \varphi \psi) , \\
&\mathcal R_{\mathcal{X}}(\varphi^{\ast}) = \sum_{\kappa = 1}^{\mx}{ \left( \int_{\Omegax}{ \varphi^{\ast} g_{\kappa} \, dx } \right) \left( \int_{\omegax}{ \lambda h_{\kappa} \, d\mu_{x} } \right) } \sum_{j = 1}^{\my}{ \left( \int_{\Omegay}{ \psi k_{j} \, dy } \right) \left( \int_{\omegay}{ \eta \ell_{j} \, d\mu_{y} } \right) } \\
& \hspace{0.75in} 
- \sum_{i = 1}^{m - 1}{ \left( \int_{\omegax}{ \lambda \lambda_{i} \, d\mu_{x} } \right) \left( \int_{\omegay}{ \eta \eta_{i} \, d\mu_{y} } \right) B(\varphi^{\ast} \psi_{i}, \varphi_{i} \psi_{i}) } .
\end{aligned}
\]
\end{formulation}

\begin{formulation}[PGD problem in parameter~$y$ for the Poisson equation]
\label{frm:poisson_pgd_y}
Assuming $\varphi$, $\lambda$, and $\eta$ to be known and taking test functions in the form $v(x,y,\mu_x,\mu_y) = \varphi(x) \psi^{\ast}(y) \lambda(\mu_{x}) \eta(\mu_{y})$, the weak formulation of the adjoint problem reads:
\[
\text{Find} \ \psi \in V_{\mathcal{Y}}\ \text{such that} \quad B_{\mathcal{Y}}(\psi, \psi^{\ast}) = \mathcal R_{\mathcal{Y}}(\psi^{\ast}), \quad \forall \psi^{\ast} \in V_{\mathcal{Y}},
\]
where:
\[
\begin{aligned}
&B_{\mathcal{Y}}(\psi, \psi^{\ast}) = \left( \int_{\omegax}{ \lambda^{2} \, d\mu_{x} } \right) \left( \int_{\omegay}{ \eta^{2} \, d\mu_{y} } \right) B(\varphi \psi^{\ast}, \varphi \psi) , \\
&\mathcal R_{\mathcal{Y}}(\psi^{\ast}) = \sum_{\kappa = 1}^{\mx}{ \left( \int_{\Omegax}{ \varphi g_{\kappa} \, dx } \right) \left( \int_{\omegax}{ \lambda h_{\kappa} \, d\mu_{x} } \right) } \sum_{j = 1}^{\my}{ \left( \int_{\Omegay}{ \psi^{\ast} k_{j} \, dy } \right) \left( \int_{\omegay}{ \eta \ell_{j} \, d\mu_{y} } \right) } \\
&\hspace{0.75in} 
- \sum_{i = 1}^{m - 1}{ \left( \int_{\omegax}{ \lambda \lambda_{i} \, d\mu_{x} } \right) \left( \int_{\omegay}{ \eta \eta_{i} \, d\mu_{y} } \right) B(\varphi_{i} \psi^{\ast}, \varphi_{i} \psi_{i}) } .
\end{aligned}
\]
\end{formulation}

\begin{remark}
In Formulation~\ref{frm:poisson_pgd_x}, the bilinear form $B_{\mathcal{X}}(\cdot,\cdot)$ results from projecting $B(\cdot,\cdot)$ onto the $x$-direction, with multiplicative weights arising from the parametric factors and the fixed $y$-mode $\psi$; analogously, in Formulation~\ref{frm:poisson_pgd_y}, $B_{\mathcal{Y}}(\cdot,\cdot)$ is determined with multiplicative weights and the fixed $x$-mode $\varphi$. Using the Finite Element Method, the construction of $B_{\mathcal{X}}$ (or $B_{\mathcal{Y}}$) reduces to assembling one-dimensional stiffness and mass matrices. The separation structure ensures that the computation of $\varphi$ or $\psi$ is carried out through a sequence of standard 1D finite element problems (or linear systems) with modified coefficients. From an algorithmic perspective, the implementation alternates between assembling $B_{\mathcal{X}}$ (fixing $\psi$, $\lambda$ and $\eta$) or $B_{\mathcal{Y}}$ (fixing $\varphi$, $\lambda$ and $\eta$), leading to a typical alternating direction scheme for PGD solvers.
\end{remark}

\begin{formulation}[PGD problem in parameter~$\mu_{x}$ for the Poisson equation]
\label{frm:poisson_pgd_mux}
Assuming $\varphi$, $\psi$, and $\eta$ to be known and taking test functions in the form~$v(x,y,\mu_x,\mu_y) = \varphi(x) \psi(y) \lambda^{\ast}(\mu_{x}) \eta(\mu_{y})$, the weak formulation of the adjoint problem reads:
\[
\text{Find}\ \lambda \in L^{2}(\omegax)\ \text{such that} \quad B_{\mathcal{X}}^{\mu}(\lambda, \lambda^{\ast}) = \mathcal R_{\mathcal{X}}^{\mu}(\lambda^{\ast}), \quad \forall \lambda^{\ast} \in L^{2}(\omegax),
\]
where:
\[
\begin{aligned}
&B_{\mathcal{X}}^{\mu}(\lambda, \lambda^{\ast}) = \left( \int_{\omegay}{ \eta^{2} \, d\mu_{y} } \right) B(\varphi \psi, \varphi \psi) \int_{\omegax}{ \lambda^{\ast} \lambda \, d\mu_{x} } , \\
&\mathcal R_{\mathcal{X}}^{\mu}(\lambda^{\ast}) = \sum_{\kappa = 1}^{\mx}{ \left( \int_{\Omegax}{ \varphi g_{\kappa} \, dx } \right) \left( \int_{\omegax}{ \lambda^{\ast} h_{\kappa} \, d\mu_{x} } \right) } \sum_{j = 1}^{\my}{ \left( \int_{\Omegay}{ \psi k_{j} \, dy } \right) \left( \int_{\omegay}{ \eta \ell_{j} \, d\mu_{y} } \right) } \\
&\hspace{0.75in} 
- \sum_{i = 1}^{m - 1}{ \left( \int_{\omegay}{ \eta \eta_{i} \, d\mu_{y} } \right) B(\varphi_{i} \psi_{i}, \varphi_{i} \psi_{i}) \int_{\omegax}{ \lambda^{\ast} \lambda_{i} \, d\mu_{x} } } .
\end{aligned}
\]
\end{formulation}

\begin{formulation}[PGD problem in parameter~$\mu_{y}$ for the Poisson equation]
\label{frm:poisson_pgd_muy}
Assuming $\varphi$, $\psi$, and $\lambda$ to be known and taking test functions in the form~$v(x,y,\mu_x,\mu_y) = \varphi(x) \psi(y) \lambda(\mu_{x}) \eta^{\ast}(\mu_{y})$, the weak formulation of the adjoint problem reads:
\[
\text{Find}\ \lambda \in L^{2}(\omegay)\ \text{such that} \quad B_{\mathcal{Y}}^{\mu}(\eta, \eta^{\ast}) = \mathcal R_{\mathcal{Y}}^{\mu}(\eta^{\ast}), \quad \forall \eta^{\ast} \in L^{2}(\omegay),
\]
where:
\[
\begin{aligned}
&B_{\mathcal{Y}}^{\mu}(\eta, \eta^{\ast}) = \left( \int_{\omegax}{ \lambda^{2} \, d\mu_{x} } \right) B(\varphi \psi, \varphi \psi) \int_{\omegay}{ \eta^{\ast} \eta \, d\mu_{y} } , \\
&\mathcal R_{\mathcal{Y}}^{\mu}(\eta^{\ast}) = \sum_{\kappa = 1}^{\mx}{ \left( \int_{\Omegax}{ \varphi g_{\kappa} \, dx } \right) \left( \int_{\omegax}{ \lambda h_{\kappa} \, d\mu_{x} } \right) } \sum_{j = 1}^{\my}{ \left( \int_{\Omegay}{ \psi k_{j} \, dy } \right) \left( \int_{\omegay}{ \eta^{\ast} \ell_{j} \, d\mu_{y} } \right) } \\
&\hspace{0.75in}
- \sum_{i = 1}^{m - 1}{ \left( \int_{\omegax}{ \lambda \lambda_{i} \, d\mu_{x} } \right) B(\varphi_{i} \psi_{i}, \varphi_{i} \psi_{i}) \int_{\omegay}{ \eta^{\ast} \eta_{i} \, d\mu_{y} } } .
\end{aligned}
\]
\end{formulation}

\begin{remark}
In Formulation~\ref{frm:poisson_pgd_mux} (resp.\ Formulation~\ref{frm:poisson_pgd_muy}), the bilinear form $B_{\mathcal{X}}^{\mu}(\cdot,\cdot)$ (resp.~$B_{\mathcal{Y}}^{\mu}(\cdot,\cdot)$) no longer involves differential operators, since Formulation~\ref{frm:pgd_wr_poisson} contains no derivatives with respect to the parameter $\mu_{x}$ (resp.~$\mu_{y}$). These formulations are solved in their strong form, leading to algebraic equations that must be evaluated for values of the parameter $\mu_{x} \in \omegax$ (resp.~$\mu_{y} \in \omegay$).
\end{remark}

Subsequently to the offline phase, estimates of~$Q_{\mu}(u) = u(\mu_{x}, \mu_{y})$, denoted here by~$\hat{u}(\mu_{x}, \mu_{y})$, can be obtained for any input data~$f : \Omega \rightarrow \mathbb{R}$ and any point $(\mu_{x}, \mu_{y}) \in \omega$. Using the linear property of the linear form~$F$, the estimates of the QoI are thus efficiently computed as:
\begin{equation}
\label{eq:poisson_qoi}
\begin{aligned}
\hat{u}(\mu_{x}, \mu_{y}) 
= F(z_{m})
&= \sum_{i = 1}^{m}{ F(\varphi_{i}  \psi_{i}) \, \lambda_{i}(\mu_{x}) \eta_{i}(\mu_{y}) } \\
&= \sum_{i = 1}^{m}{ \left( \int_{\Omega}{ f(x,y) \, \varphi_{i}(x) \psi_{i}(y) \, dxdy } \right) \lambda_{i}(\mu_{x}) \eta_{i}(\mu_{y}) } .
\end{aligned}
\end{equation}
The contraction of such an expression is fast, especially when the rank~$m$ of the PGD approximation is low. This task is commonly referred to as the \emph{online phase}.

\subsection{Plane-Stress Elasticity}

As above, we assume that the kernel function $k_\epsilon$ is given in terms of the Gaussian function, for $x\in \Omega \subset \mathbb R^2$ and $\mu \in \gamma \subset \Gamman$, and takes the form
\[
k_{\epsilon}(x-\mu) = \frac{1}{\sqrt{2 \pi} \epsilon} \, \exp \left( -\frac{\| x - \mu \|^{2}}{2 \epsilon^{2}} \right) ,
\]
where $\|\cdot\|$ is the Euclidean norm in $\mathbb R^2$. We also assume that it can be represented as the separated expansion:
\begin{equation}
\label{eq:ps_elas_gk}
k_{\epsilon}(x-\mu) \simeq \sum_{\kappa = 1}^{n}{ h_{\kappa}(x) \, k_{\kappa}(\mu) } .
\end{equation}

\begin{formulation}[Weighted residual formulation of the parameterized adjoint problem for plane-stress elasticity]
\label{frm:param_wr_ps_elas}
\[
\text{Find}\ z_{\mu} \in U \times L^{2}(\gamma)\ \text{such that}\quad \int_{\gamma}{ A(v, z_{\mu}) \, d\mu} = \int_{\gamma}{ J_{\mu}(v) \, d\mu}, \quad \forall v \in U \times L^{2}(\gamma). 
\]
\end{formulation}

\noindent 
We now seek an approximation of the solution to the adjoint problem as:
\[
z_{m}(x, \mu) 
= \sum_{i = 1}^{m} \varphi_{i}(x) \lambda_{i}(\mu) 
= \bigg( \sum_{i = 1}^{m-1} \varphi_{i}(x) \lambda_{i}(\mu)  \bigg)
+ \varphi(x) \lambda(\mu).
\]
The fixed-point iteration is very similar to that illustrated in Figure~\ref{fig:flowchart_fixedpoint}, but this time with two subproblems to solve instead of four as we suppose here that we cannot separate the spatial mode $\varphi_{i}(x)$, nor the parameter mode $\lambda_{i}(\mu)$, with respect to the directions $x$ and $y$ due to the complexity of the geometry. The sequence $\left( \varphi^{(k)} \right)_{k \in \mathbb{N}}$ is also accelerated by means of the Aitken's $\Delta^{2}$ process.

\begin{formulation}[PGD problem in parameter~$x$ for plane-stress elasticity]
\label{frm:ps_elas_x}
Assuming $\lambda$ to be known and taking test functions in the form $v (x,\mu) = \varphi^{\ast}(x) \lambda(\mu)$, the weak formulation of the adjoint problem reads:
\[
\text{Find}\ \varphi \in U \ \text{such that} \quad A_{\Omega}(\varphi, \varphi^{\ast}) = \mathcal R_{\Omega}(\varphi^{\ast}), \quad \forall \varphi^{\ast} \in U,
\]
where:
\[
\begin{aligned}
&A_{\Omega}(\varphi, \varphi^{\ast}) = \left( \int_{\gamma}{ \lambda^{2} \, d\mu } \right) A(\varphi, \varphi^{\ast}) , \\
&\mathcal R_{\Omega}(\varphi^{\ast}) = \sum_{\kappa = 1}^{n}{ \left( \int_{\Gamman}{ \varphi^{\ast} \cdot n \, h_{\kappa} \, ds } \right) \left( \int_{\gamma}{ \lambda k_{\kappa} \, d\mu } \right) } - \sum_{i = 1}^{m - 1}{ \left( \int_{\gamma}{ \lambda \lambda_{i} \, d\mu } \right) A(\varphi_{i}, \varphi^{\ast}) } .
\end{aligned}
\]
\end{formulation}

\begin{formulation}[PGD problem in parameter~$\mu$ for plane-stress elasticity]
\label{frm:ps_elas_mu}
Assuming $\varphi$ to be known and taking test functions in the form $v (x,\mu) = \varphi(x) \lambda^{\ast}(\mu)$, the weak formulation of the adjoint problem reads:
\[
\text{Find}\ \lambda \in L^{2}(\gamma) \ \text{such that} \quad A_{\gamma}^{\mu}(\lambda, \lambda^{\ast}) = \mathcal R_{\gamma}^{\mu}(\lambda^{\ast}), \quad \forall \lambda^{\ast} \in L^{2}(\gamma),
\]
where:
\[
\begin{aligned}
&A_{\gamma}^{\mu}(\lambda, \lambda^{\ast}) = A(\varphi, \varphi) \int_{\gamma}{ \lambda^{\ast} \lambda \, d\mu } , \\
&\mathcal R_{\gamma}^{\mu}(\lambda^{\ast}) = \sum_{\kappa = 1}^{n}{ \left( \int_{\Gamman}{ \varphi \cdot n \, h_{\kappa} \, ds } \right) \left( \int_{\gamma}{ \lambda^{\ast} k_{\kappa} \, d\mu } \right) } - \sum_{i = 1}^{m - 1}{ A(\varphi_{i}, \varphi) \int_{\gamma}{ \lambda^{\ast} \lambda_{i} \, d\mu } } .
\end{aligned}
\]
\end{formulation}

\begin{remark}
Similarly to the Poisson equation, the PGD algorithm alternates here between Formulation~\ref{frm:ps_elas_x}, which solves a finite element problem for $\varphi$, and Formulation~\ref{frm:ps_elas_mu}, which consists in a system of algebraic equations for $\lambda$, for values of $\mu \in \gamma$.
\end{remark}

Since $L$ is a linear form, the estimates of the QoI, for any input data $f : \Omega \rightarrow \mathbb{R}^{2}$ and $g : \Gamman \rightarrow \mathbb{R}^{2}$ and for any $\mu \in \gamma \subset \Gamman$, can thus be computed in the \emph{online phase} as:
\begin{equation}
\label{eq:ps_elas_qoi}
\hat{u}(\mu) \cdot n = L(z_{m})	
= \sum_{i = 1}^{m}{ L(\varphi_{i}) \lambda_{i}(\mu) } = \sum_{i = 1}^{m}{ \left( \int_{\Omega}{ f \cdot \varphi_{i} \, dx + \int_{\Gamman} g \cdot \varphi_{i} \, ds } \right) \lambda_{i}(\mu) } .
\end{equation}

\section{Numerical examples}
\label{sec:num_ex}

We now introduce two examples based on the model problems presented above. The first one will deal with the Poisson problem and will serve as a proof-of-concept of the proposed methodology. The second example is concerned with the design of a structural bracket using plane-stress elasticity to illustrate the performance of the methodology on an application of engineering interest. 

\subsection{Example with the Poisson Equation}

We consider the unit square~$\Omega = \Omegax \times \Omegay$ with $\Omegax = \Omegay = (0,1)$. The subdomain of interest is  defined as the tensor-product~$\omega = \omegax \times \omegay$ with $\omegax = \omegay = (0.2, 0.8)$, see Figure~\ref{fig:poisson_omega}. We are interested in solving the primal problem~\eqref{eq:poisson_eq} for the three distinct source terms reported in Table~\ref{tab:tc1}, the objective being to assess the robustness of the method for various inputs. For given $\mu \in \omega$, we recall the adjoint problem:
\[
\begin{aligned}
- \Delta z_{\mu} = \bar{k}_\epsilon (x-\mu_{x}), 
& \quad \forall x \in \Omega, \\
z_{\mu} = 0, 
& \quad \forall x \in \partial \Omega.
\end{aligned}
\]
where the kernel function $\bar{k}_\epsilon$ is a separated approximation~\eqref{eq:poisson_gk} of the Gaussian kernel:
\[
k_\epsilon (x-\mu_{x},y-\mu_{y}) = \frac{1}{2 \pi \epsilon^{2}} \, \exp \left( -\frac{(x - \mu_{x})^{2}}{2 \epsilon^{2}} \right) \, \exp \left( -\frac{(y - \mu_{y})^{2}}{2 \epsilon^{2}} \right)
\]
The domain $\Omega$ is partitioned into a structured mesh of bilinear quadrilateral elements made of 249,001 elements resulting in a total of 250,000 degrees of freedom. The standard deviation in the Gaussian kernel is set to $\epsilon = 4 \times 10^{-3}$.

\begin{table}[tbh]
\caption{Source terms for the Poisson example}
\label{tab:tc1}
\begin{tabular*}{0.5\textwidth}{@{\extracolsep\fill}ll}
\toprule%
{\bfseries Source term} & {\bfseries Expression} \\ \midrule %
$f_{1}$ & $1000$ \\ \midrule %
$f_{2}$ & $1000 \, x y^2$ \\ \midrule %
$f_{3}$ & $1000 \cos(6 \pi x) \sin(2 \pi y)$ \\
\botrule
\end{tabular*}
\end{table}

We assess the accuracy of the PGD approximations with respect to the finite element solutions $u_h$ solving the full model. More specifically, we evaluate the relative errors in the QoI estimates $\hat{u}$~\eqref{eq:poisson_qoi} as follows:
\[
\aerro = 
\frac{\|\hat{u}-Q_{\mu}(\ufem)\|_{L^{2}(\omega)}}{\| Q_{\mu}(\ufem)\|_{L^{2}(\omega)}}. 
\]
In practice, the reference $Q_{\mu}(\ufem)$ is evaluated directly from the finite element solution $\ufem$, i.e., $Q_{\mu}(\ufem) = \ufem$ in $\omega$, without using the kernel representation. As a result, the error $\aerro$ includes both the model reduction error associated with the PGD approximation and the additional approximation errors introduced by the kernel-based representation of the QoI.

\begin{figure}[tbh]
	\centering
	\includegraphics[width=.95\textwidth]{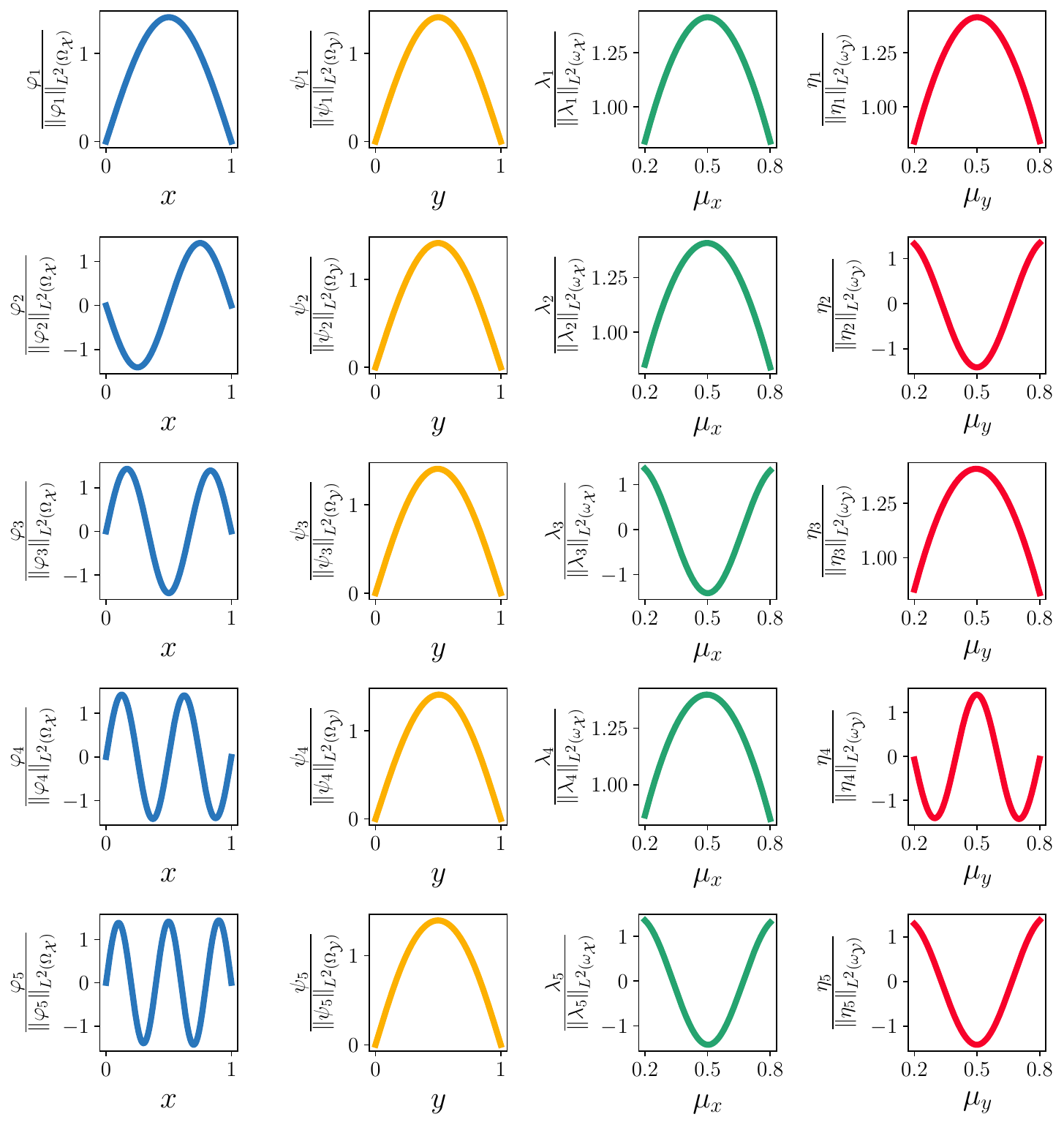}
	\caption{Poisson equation -- First five PGD modes of the Parameterized Adjoint Formulation~\ref{frm:pgd_wr_poisson}.}
    \label{fig:poisson_modes}
\end{figure}

\begin{figure}[tbh]
	\centering
	\includegraphics[width=\textwidth]{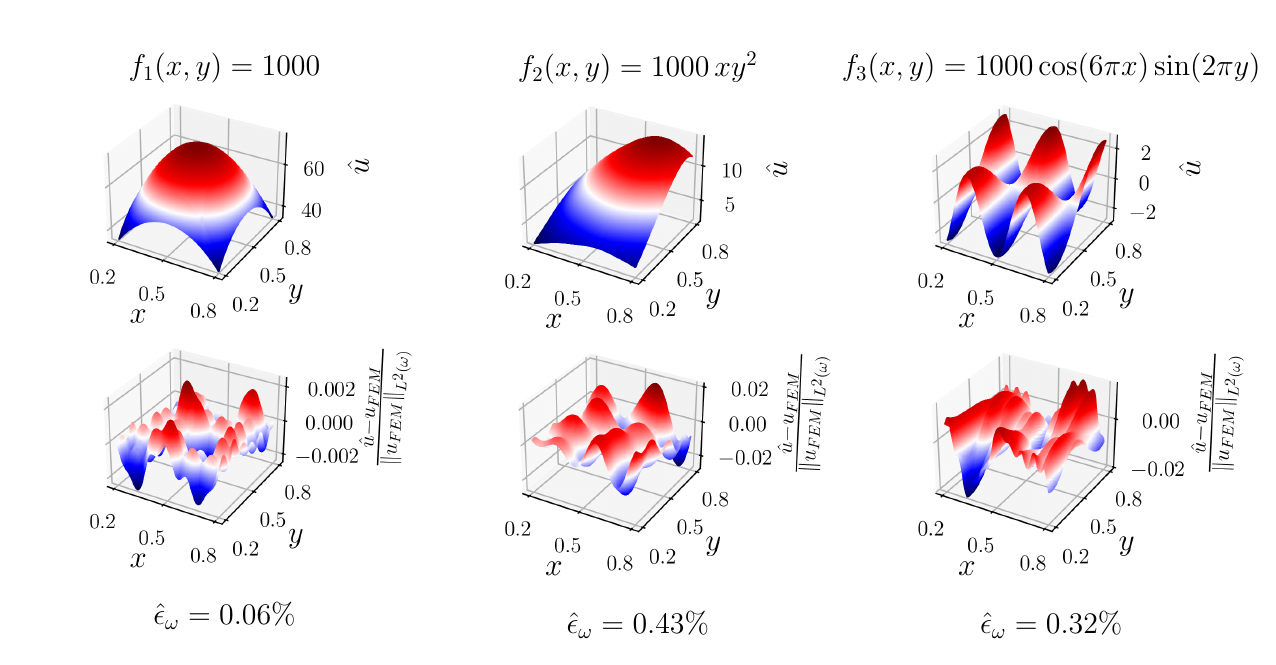}
	\caption{Poisson equation -- Estimation of the QoI using the PGD surrogate model obtained from the Parameterized Adjoint Formulation~\ref{frm:pgd_wr_poisson} with 50 modes and associated errors.}
    \label{fig:pgd_vs_fem_poisson}
\end{figure}

Figures~\ref{fig:poisson_modes} and~\ref{fig:pgd_vs_fem_poisson} show results of the PGD-based surrogate model using the adjoint formulation of the Poisson equation. The first five PGD modes shown in Figure~\ref{fig:poisson_modes} capture the dominant parametric dependence of the adjoint solution, providing a compact reduced basis for the surrogate model. The QoI estimates obtained with the PGD surrogate using the first $50$ modes match the FEM-based reference values with an error below $1\%$ for the three considered load configurations, as shown in Figure~\ref{fig:pgd_vs_fem_poisson}.

From a computational standpoint, the benefits of the surrogate approach are tangible on this toy problem, albeit on a limited scale. The resolution of the FEM problem requires about 31 seconds, whereas the PGD offline stage is completed in about 10 seconds. Once constructed, the cost per evaluation is significantly reduced: a single FEM solve, i.e., the cost of the forward and backward substitutions, takes around 50 ms per load case, while the corresponding PGD evaluation requires roughly 1 ms. These gains indicate the expected scaling advantage of the approach, with substantially larger savings anticipated for problems of engineering relevance, with larger scales and involving many load cases.

This level of accuracy confirms that the proposed methodology, replacing repeated primal solves by a surrogate model for the adjoint problem, can provide reliable QoI predictions at a fraction of the computational cost, thereby validating its suitability for scenarios requiring evaluations under multiple loading conditions. 

\subsection{Example with Plane-Stress Elasticity}

\begin{figure}[tb]
\centering
\includegraphics[width=.45\textwidth]{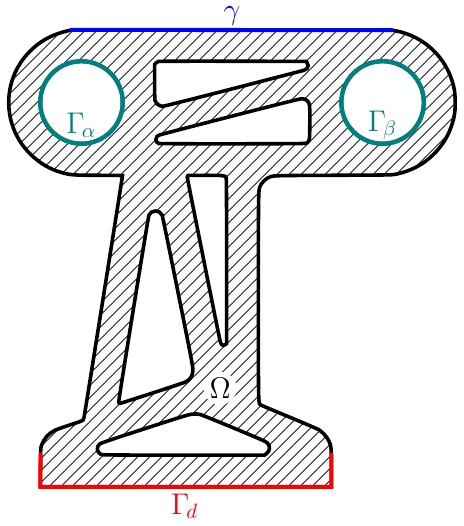}
\hspace{0.05\textwidth}
\includegraphics[width=.40\textwidth]{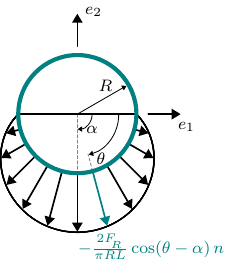}
\caption{Plane-stress elasticity – Left: Schematic of the 2D computational domain $\Omega$, Neumann boundaries $\Gammal$ and $\Gammar$ for the primal problem, Neumann boundary $\gamma$ for the adjoint problem, and Dirichlet boundary $\Gammad$; Right: Schematic of the bearing load with an offset $\alpha=- \pi / 2$ pointing in the opposite direction of the unit outward normal vector $n$, which points here towards the center of the bore, that is, $n = - (\cos(\theta),\sin(\theta))$.}
\label{fig:bracket_drawing}
\end{figure}

This example deals with the validation of a structural bracket, shown in Figure~\ref{fig:bracket_drawing}, subjected to a wide range of load configurations, which is representative of a typical scenario in design optimization. The goal is to efficiently estimate a QoI under varying loading conditions. We consider the following plane-stress problem:
\[
\begin{aligned}
- \nabla \cdot \sigma(u) = 0, 
&\quad \text{in}\ \Omega, \\
\sigma(u) = \mathbf{C} : \varepsilon(u), 
&\quad \text{in}\ \Omega, \\
\varepsilon(u) = \frac{1}{2} \left(\nabla u + (\nabla u)^{\top} \right), 
&\quad \text{in}\ \Omega, \\
u = 0, 
&\quad \text{on}\ \Gammad, \\
\sigma(u) \cdot n = g_{\alpha} \, n, 
&\quad \text{on}\ \Gammal, \\
\sigma(u) \cdot n = g_{\beta} \, n, 
&\quad \text{on}\ \Gammar, \\
\sigma(u) \cdot n = 0, 
&\quad \text{on}\ \partial \Omega \setminus ( \Gammad \cup \Gammal \cup \Gammar ),
\end{aligned}
\]
where the Neumann boundary conditions on the portions $\Gammal$ and $\Gammar$ of the boundary depend on angles $(\alpha, \beta) \in \mathcal{D}^2$, with $\mathcal{D} = [0, 2 \pi)$, which determine the loading zone on the bearing seat, as shown in Figure~\ref{fig:bracket_drawing}.
More specifically, $\Gammal$ and $\Gammar$ are subjected to the bearing loads:
\[
\label{eq:bearingloads}
\begin{aligned}
g_{\alpha}(\theta) &= - \frac{2 \Fr}{\pi R L} \cos(\theta - \alpha) \, \ind{\alpha}(\theta) , \\
g_{\beta}(\theta) &= - \frac{2 \Fr}{\pi R L} \cos(\theta - \beta) \, \ind{\beta}(\theta) ,
\end{aligned}
\]
where $\Fr$ is the magnitude of the resultant of the radial load, $R$ and $L$ are the radius and width of the bearing seat, respectively, and $\ind{\alpha}$ denotes the indicator function with respect to parameter $\alpha$ given by:
\[
\ind{\alpha}(\theta) =  \left\{ \begin{aligned}
1, & \quad \text{if} \ \alpha - \frac{\pi}{2} < \theta < \alpha + \frac{\pi}{2}, \\
0, & \quad \text{otherwise} .
\end{aligned} \right.
\]
The indicator function $\ind{\beta}$ with respect to $\beta$ is defined in a similar manner. Later on, the load parameters $\alpha$ and $\beta$ will be sampled on an evenly spaced grid over the interval $\mathcal{D} = [0, 2\pi)$. In this work, both grids consist of $\na = \nb = 360$ points, that is:
\[
\alpha_{i} = \frac{2\pi (i-1)}{360}, \quad i = 1, \ldots, 360,
\]
and similarly for $\beta$.

We suppose here is that we interested in the normal displacement along the top boundary $\gamma$ of the structural bracket. The adjoint problem then reads:
\[
\begin{aligned}
- \nabla \cdot \sigma(z_{\mu}) = 0, 
&\quad \text{in}\ \Omega, \\
\sigma(z_{\mu}) = \mathbf{C} : \varepsilon(z_{\mu}), 
&\quad \text{in}\ \Omega, \\
\varepsilon(z_{\mu}) = \frac{1}{2} \left(\nabla z_{\mu} + (\nabla z_{\mu})^{\top} \right), 
&\quad \text{in}\ \Omega, \\
z_{\mu} = 0, 
&\quad \text{on}\ \Gammad, \\
\sigma(z_{\mu}) \cdot n 
= \frac{1}{\sqrt{2 \pi} \epsilon} \, 
\exp \left( -\frac{\| x - \mu \|^{2}}{2 \epsilon^{2}} \right) n, 
&\quad \text{on}\ \gamma, \\
\sigma(z_{\mu}) \cdot n = 0, 
&\quad \text{on}\ \partial \Omega \setminus (\Gammad\cup\gamma) .
\end{aligned}
\]
where the standard deviation in the Gaussian kernel is set to $\epsilon = 5 \times 10^{-2}$. Moreover, the Lam\'{e} parameters $\lambda$ and $\mu$ in Hooke's law~\eqref{eq:HookesLaw} are evaluated here from the Young's Modulus and Poisson ratio as (see Table~\ref{tab:tc2}):
\[
\lambda = \frac{E \nu}{1 - \nu^{2}} 
\quad \text{and} \quad 
\mu = \frac{E}{2 (1 + \nu)}.
\] 

\begin{table}[tbh]
\caption{Parameters for the plane-stress elasticity example}
\label{tab:tc2}
\begin{tabular*}{0.6\textwidth}{@{\extracolsep\fill}ll}
\toprule%
{\bfseries Parameter} & {\bfseries Value} \\ \midrule %
Young's Modulus $E$ & $70 \times 10^{3}$ MPa \\ \midrule %
Poisson ratio $\nu$ & 0.32 \\ \midrule %
Radius $R$ & 20 mm \\ \midrule %
Magnitude $\Fr$ & 500 N \\ \midrule %
\footnotemark[1]Width $L$ & 1 SI \\ 
\botrule
\end{tabular*}
\footnotetext[1]{Introduced for the purpose of unit consistency.}
\end{table}

The domain $\Omega$ is partitioned into a mesh of Lagrange quadrilateral elements of degree one, consisting of 993,893 elements and a total of 2,002,946 degrees of freedom. The mesh is chosen fine enough in order to neglect the discretization errors. We note that we will use the same mesh to compute both the FEM solutions $\ufem$ of the full model and the modes of the PGD approximations of the reduced model.

We shall assess the performance of the proposed methodology in terms of the accuracy of the approximations and the time efficiency:
\begin{enumerate}
\item 
The accuracy of the PGD approximations will be measured in term of the relative error $\aerrg(\alpha,\beta)$ and root mean square relative error $\hat{\eta}_{\gamma}$ of the quantity of interest $\hat{u} = L(z_{m})$, computed from Eq.~\eqref{eq:ps_elas_qoi}, with respect to that computed from the finite element solutions $\ufem$, i.e.\ $J_{\mu}(\ufem)$, of the full-order model (FOM), for $(\alpha,\beta) \in \mathcal{D}^2$:
\[
\begin{aligned}
&\aerrg(\alpha,\beta) 
= \frac{ \| 
\hat{u}(\alpha,\beta) - 
J_{\mu}(\ufem(\alpha,\beta)) \|_{L^{2}(\gamma)} }{ \| 
J_{\mu}(\ufem(\alpha,\beta)) \|_{L^{2}(\gamma)} }, 
\\[.1in]
&\hat{\eta}_{\gamma} 
= \sqrt{ \frac{1}{\na \nb} 
\sum_{i = 1}^{\na} 
\sum_{j = 1}^{\nb} 
\big( \aerrg(\alpha_{i}, \beta_{j}) \big)^{2} }.
\end{aligned}
\]
For comparison purposes, we will also evaluate the quantity of interest $J_{\mu}(u_{m})$ by considering the PGD approximation $u_{m}$~\eqref{eq:pgd_ps_elas} of the primal problem, as shown in Appendix~\ref{apx:primal_pgd}. We will compute the relative error $\perrg(\alpha,\beta)$ and root mean square relative error $\eta_{\gamma}$ in $J_{\mu}(u_{m})$ with respect to that obtained using the FOM: 
\[
\begin{aligned}
&\perrg(\alpha,\beta) 
= \frac{ \| 
J_{\mu}(u_{m}(\alpha,\beta)) - 
J_{\mu}(\ufem(\alpha,\beta)) \|_{L^{2}(\gamma)} }{ \| 
J_{\mu}(\ufem(\alpha,\beta)) \|_{L^{2}(\gamma)} }, 
\\[.1in]
&\eta_{\gamma} 
= \sqrt{ \frac{1}{\na \nb} 
\sum_{i = 1}^{\na}
\sum_{j = 1}^{\nb} 
\big( \perrg(\alpha_{i},\beta_{j}) \big)^{2} }.
\end{aligned}
\]
As for the Poisson example, the QoI $J_{\mu}$ are evaluated directly from the approximate displacement solutions, i.e., $J_{\mu}(u) = u$ on $\gamma$, without using the kernel representation. Thus, the reported errors include both the PGD model reduction error and the approximation errors associated with the kernel-based QoI.

\item 
Computational efficiency of the proposed adjoint-based reduced-order model will be assessed with respect to the FOM and the primal-based PGD ROM by measuring wall-clock times when evaluating the structural response for multiple load configurations. In the case of the FOM solved by the FEM, the total time includes the Cholesky factorization of the stiffness matrix (performed once) and the forward/backward substitutions required for each right-hand side. Since the model is linear, the factorization can be reused for all load cases, and, by the superposition principle, only $\nl = \na + \nb$ substitutions are required instead of $\na \times \nb$. In the case of the ROM, the total time amounts to the \emph{training phase}, during which the surrogate model is built, and the \emph{online phase}, during which the surrogate model is queried for any load configuration. 
Moreover, we compare the adjoint-based ROM with the FOM in terms of the operation counts:
\[
\begin{aligned}
&\cfom = \cfac + \nl \, \csub, \\
&\crom = \coff + \nl \, \con,
\end{aligned}
\]
where the different terms are defined in Table~\ref{tab:costs}. In the present work, the online evaluation cost $\ton$ is usually negligible, so the operation count for the ROM is approximately the same as that of the training phase given by:
\[
\crom \approx \coff 
= \cfac + \csub \sum_{i=1}^{m}{ k_{i} }
= \cfac + m \,\bar{k} \,\csub ,
\]
where $m$ is the rank of the PGD approximation, $k_{i}$ denotes the number of fixed-point iterations per mode, and $\bar{k}$ the average number of substitutions per mode. Therefore, the proposed method is expected to be computationally advantageous whenever
$m \, \bar{k} \ll \nl$.

\begin{table}[tbh]
\caption{Summary of operation counts in the FOM and ROM}
\label{tab:costs}
\begin{tabular*}{0.78\textwidth}{@{\extracolsep\fill}ll}
\toprule
{\bfseries Symbol} & {\bfseries Description} \\
\midrule
$\cfac$ & Operation count of the Cholesky factorization\footnotemark[1], i.e.~$K = L L^{\top}$ \\
$\csub$ & Cost of one forward/backward substitution\footnotemark[2] \\
$\coff$ & Training cost of the ROM \\
$\con$  & Online evaluation cost of the ROM (negligible) \\
$\nl$   & Number of load cases \\
\botrule
\end{tabular*}
\footnotetext[1]{$\cfac \approx O(n^{3/2})$ for two-dimensional finite element graphs~\cite{gilbert1986analysis}, where $n$ is the number of degrees of freedom, i.e.~$K \in \mathbb{R}^{n \times n}$.}
\footnotetext[2]{$\csub \approx O(|L|)$, where $|L|$ is the number of nonzero entries in the Cholesky factor.}
\end{table}

All computations were run on a computer with the following configuration:
\begin{itemize}
\item CPU: AMD Ryzen 7 PRO 4750U @ 1.7 GHz per core;
\item RAM: 38 GB;
\item OS: Arch Linux 6.15.3.
\end{itemize}
The code was written in Python 3.13.3 with NumPy 2.3.0~\cite{harris} and SciPy 1.15.3~\cite{scipy}. \code{scikit-sparse 0.4.16} is imported to leverage SuiteSparse's CHOLMOD library for sparse Cholesky decomposition~\cite{chen}. \code{opt\_einsum 3.4.0} is leveraged to optimize tensor contractions~\cite{smith}. Meshes are generated with Gmsh~\cite{geuzaine2009gmsh, remacle2012blossom}.
\end{enumerate}

\begin{figure}[hp]
\centering
\includegraphics[width=.75\textwidth]{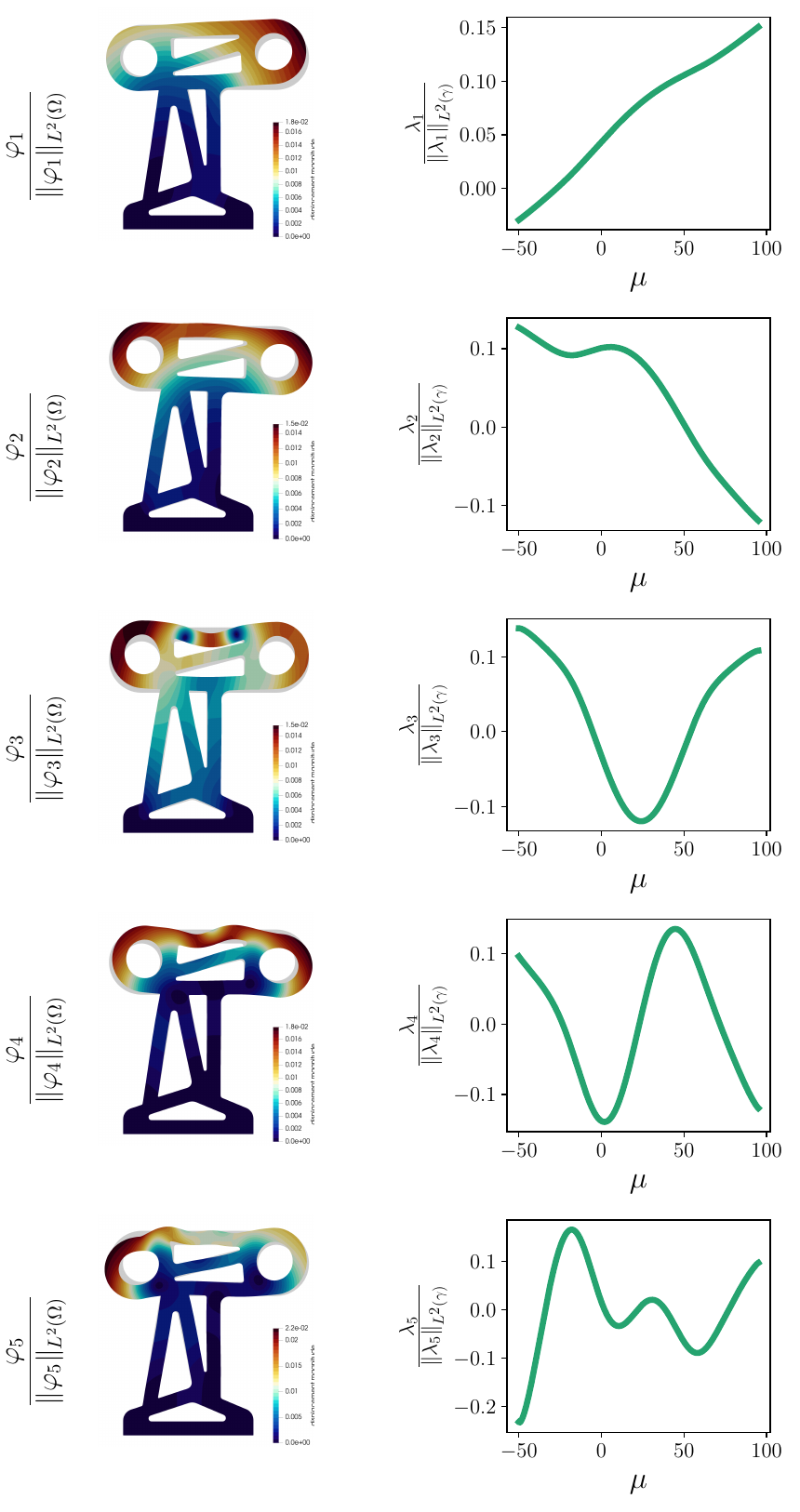}
\caption{Plane-stress elasticity -- First five PGD modes for the parameterized adjoint formulation (Formulation~\ref{frm:param_wr_ps_elas}).}
\label{fig:ps_elas_adj_modes}
\end{figure}

\begin{figure}[hp]
\centering
\includegraphics[width=.98\textwidth]{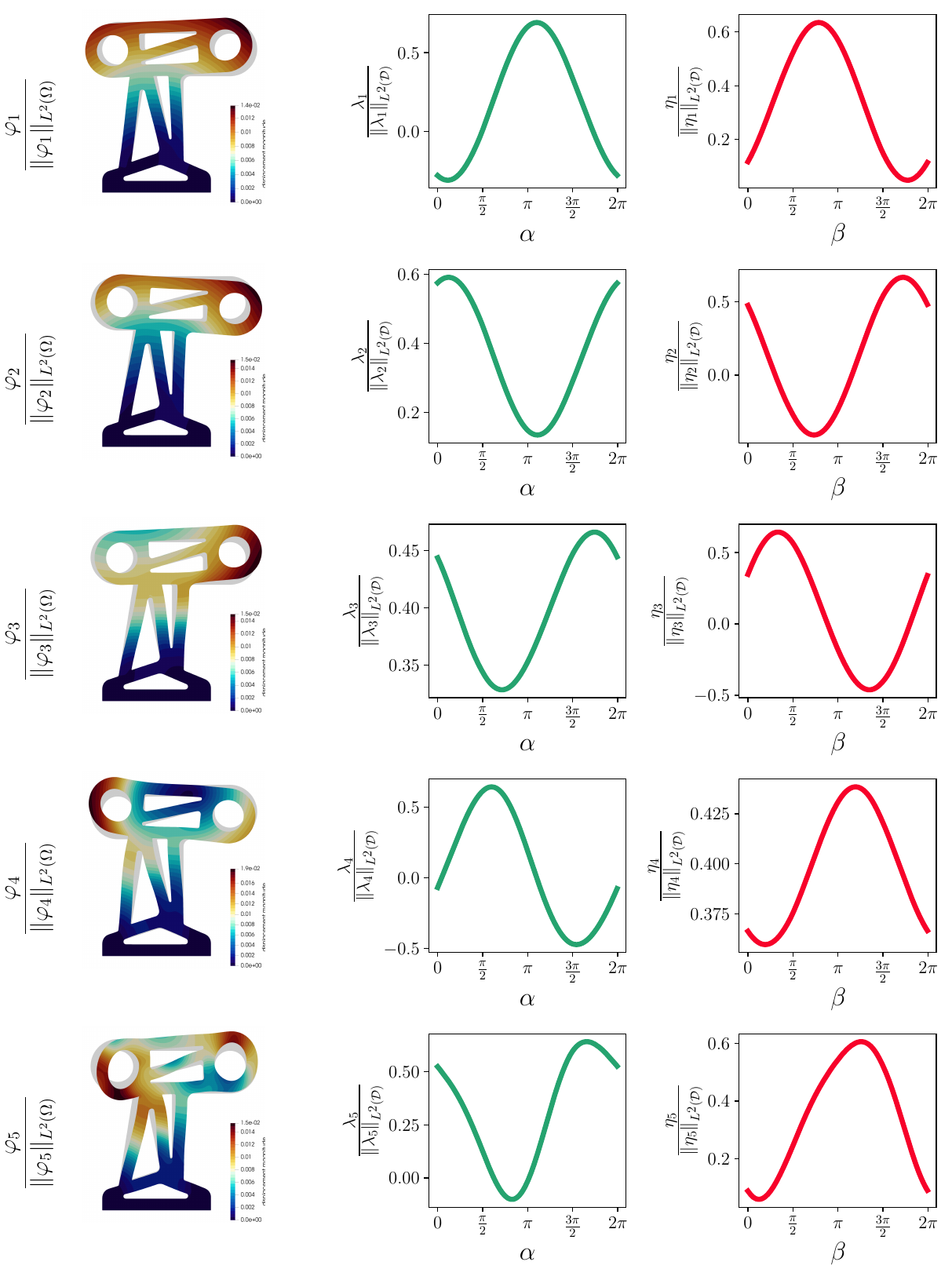}
\caption{Plane-stress elasticity -- First five PGD modes for the parameterized primal formulation (Formulation~\ref{frm:param_wr_ps_elas_pri}).}
\label{fig:ps_elas_pri_modes}
\end{figure}

\begin{figure}[ht]
\centering
\includegraphics[width=.75\textwidth]{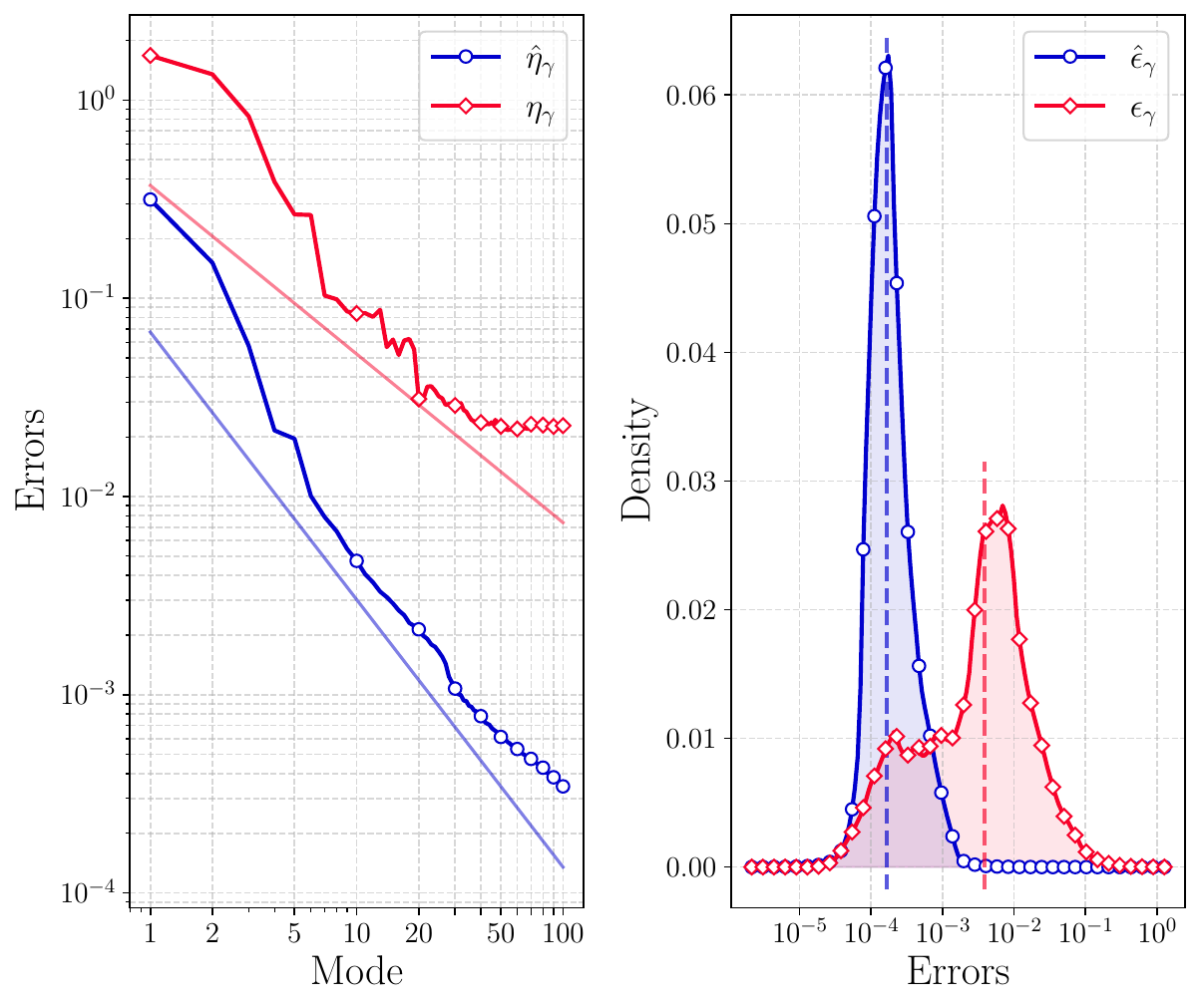}
\caption{Plane-stress elasticity -- Left: Decay of the root mean square error for both the primal and adjoint ROMs for 100 modes; Right: Density plots of the errors for both the primal and adjoint ROMs for each load configuration for 100 modes (with the median errors indicated by the dashed vertical lines).}
\label{fig:rms_adj_prim}
\end{figure}

Figures~\ref{fig:ps_elas_adj_modes} and~\ref{fig:ps_elas_pri_modes} show the first five PGD modes computed using the parameterized adjoint and primal formulations related to the plane-stress elasticity problem, respectively. While the primal modes describe the main deformation patterns of the structure under varying load configurations, the adjoint modes capture the dominant parametric behavior of the sensitivity of the quantity of interest.

Figure~\ref{fig:rms_adj_prim} quantifies the approximation errors over 100 modes, for the purpose of illustrating the asymptotic convergence trends. For the adjoint ROM, the root mean square error decays rapidly at first and then steadily decreases, exhibiting a high convergence rate for the quantity of interest. In contrast, the primal ROM shows a slower decay before plateauing at the value $\eta_\gamma \approx 2\times 10^{-2}$. One possible explanation for this slower convergence is that each primal PGD enrichment involves the product of three separable modes, whereas the adjoint-based PGD enrichment relies on only two, leading to a more complex approximation space and reduced efficiency per added mode. The density plots in Figure~\ref{fig:rms_adj_prim} further highlight the advantage of the adjoint ROM: the error distribution is narrower and the median error is one order of magnitude lower compared to the primal ROM, confirming its superior performance for QoI estimation.

\begin{table}[tbh]
\caption{Summary of the elapsed time by task}
\label{tab:time}
\begin{tabular*}{.9\textwidth}{@{\extracolsep\fill}lll}
\toprule
{\bfseries Symbol} & {\bfseries Description} & {\bfseries Wall-clock time (s)} \\
\midrule
$\tfac$ & Cholesky factorization & 33 \\
$\nl \, \tsub$ & Forward/backward substitutions & 278 \\
$\toff = \tfac + m \bar{k} \, \tsub$ & Training of the ROM & 56 \\
$\nl \, \ton$  & Online evaluation of the ROM & $5 \times 10^{-3}$ \\
\botrule
\end{tabular*}
\end{table}

The wall-clock time results reported in Table~\ref{tab:time} are obtained for 10 modes, for which the adjoint PGD commits an error below 1\% with respect to the QoI. It highlights the substantially higher efficiency of the adjoint PGD approach compared to the FOM strategy. The full FEM computation requires 311 seconds in total, including 278 seconds spent on forward and backward substitutions for all load cases. In comparison, the adjoint PGD approach incurs an offline cost of 56 seconds and an online evaluation time of only 5~ms for all loads, demonstrating its superior time efficiency.

\begin{figure}[ht]
\centering
\includegraphics[width=.95\textwidth]{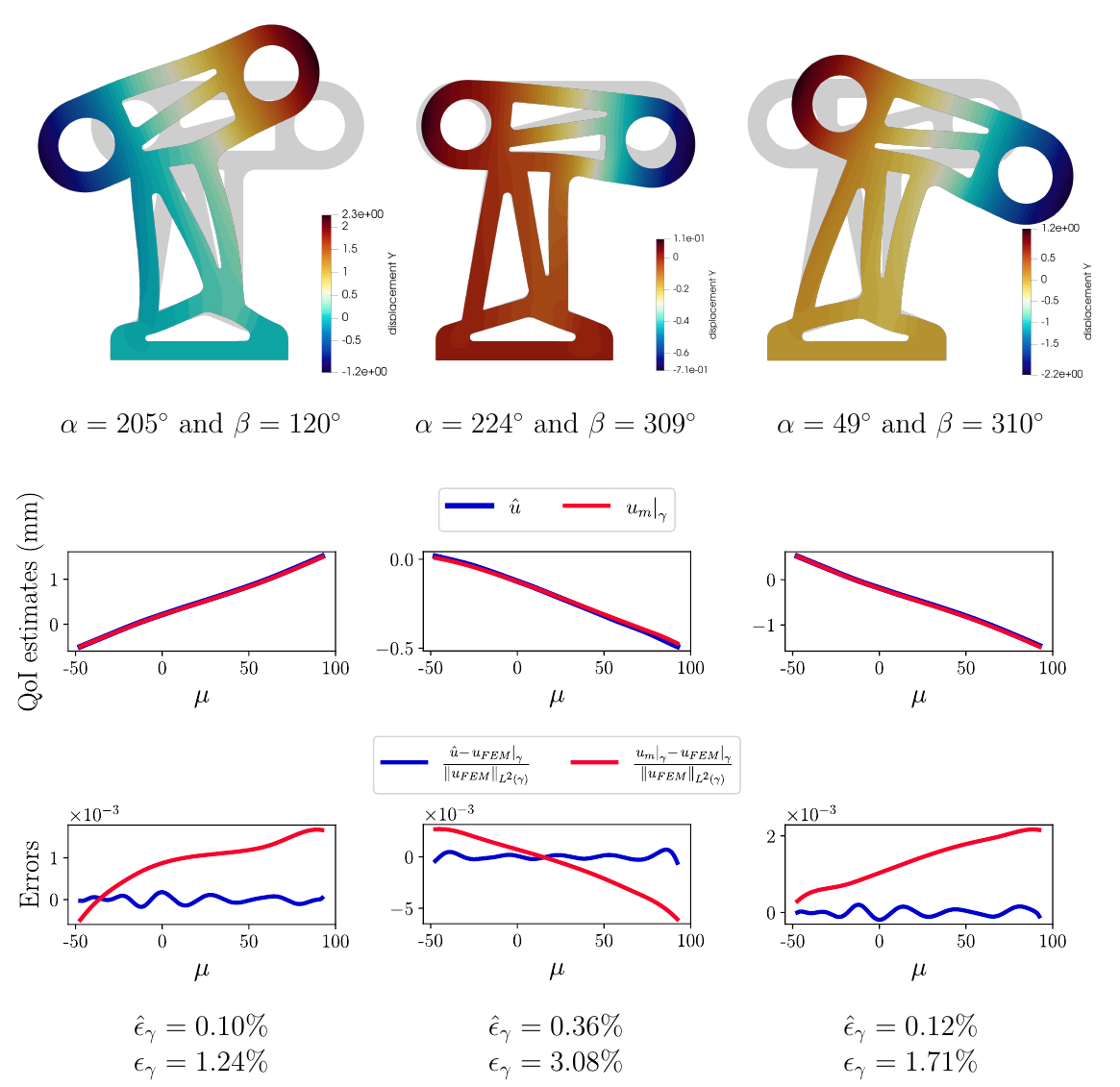}
\caption{Plane-stress elasticity -- QoI estimates over $\gamma$ for three load configurations, using primal and adjoint ROMs with 10 modes, and associated errors (full-field deformed shapes, obtained by FEM, are shown for visualization purposes).}
\label{fig:deformed_bracket}
\end{figure}

Figure~\ref{fig:deformed_bracket} illustrates the deformed shapes for three load configurations together with the associated QoI. For the same number of modes, the adjoint PGD yields QoI errors that are approximately one order of magnitude smaller than those obtained with the primal PGD, highlighting the improved accuracy of the adjoint-based approach for QoI estimation.

\begin{figure}[ht]
\centering
\includegraphics[width=.7\textwidth]{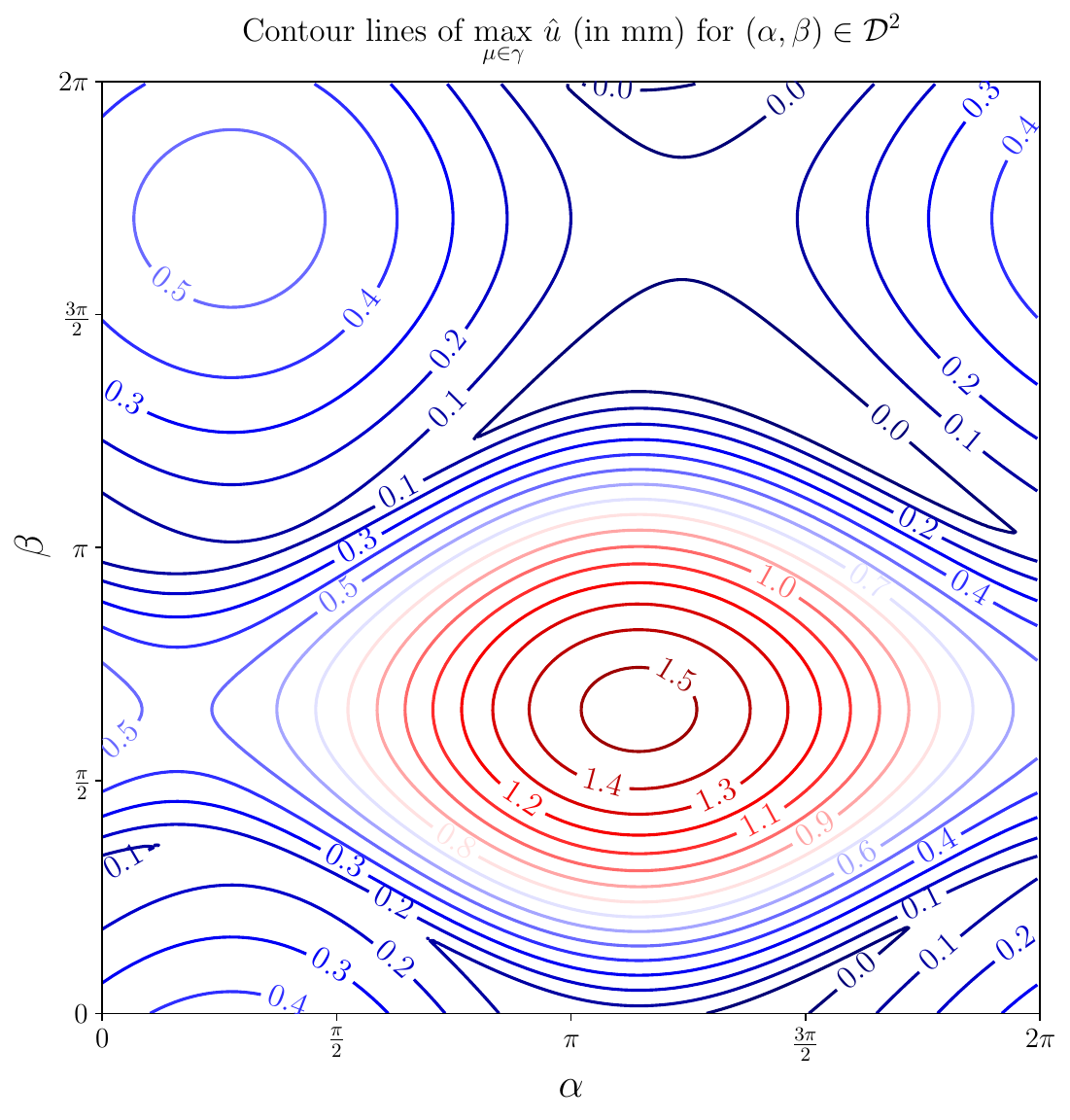}
\caption{Plane-stress elasticity -- Virtual chart generated by the adjoint surrogate model showing the maximum normal displacement on $\gamma$, for all load configurations $(\alpha, \beta) \in \mathcal{D}^{2}$.}
\label{fig:contours}
\end{figure}

Figure~\ref{fig:contours} presents a virtual chart of the maximum normal displacement along the boundary $\gamma$, obtained using the adjoint surrogate model. Fast QoI estimation enables rapid generation of such visualizations, providing timely insights into structural behavior for supporting informed decision-making. This capability would be particularly useful in generative design workflows, where it allows designers to quickly validate or invalidate design iterations based on their structural performance.

\subsection{Further Discussion}

Although not as efficient in the presented use case, the primal ROM can still provide useful reduced-order approximations of the full field solution. However, the approach is less generic than the adjoint-based strategy for linear problems. In particular, it relies on the ability to parameterize all relevant external loads. In many practical scenarios, such as industrial applications with complex, non-parametric loadings, constructing a suitable parametric representation may be challenging if not impossible. On the other hand, the adjoint-based PGD focuses directly on the QoI, decouples the surrogate model from the specifics of the loading configuration, and allows one to  efficiently evaluate the QoI for multiple load scenarios without having to recompute the primal solution.

On another note, it is worth mentioning that the SVDs performed to approximate the kernel functions~\eqref{eq:poisson_gk} and~\eqref{eq:ps_elas_gk} were not truncated, as this choice maximizes accuracy while having a negligible impact on efficiency thanks to the low computational cost of evaluating one-dimensional integrals. In fact, the separation of the Gaussian kernels with respect to their spatial parameters ($x$ or $y$) and expected value parameters ($\mu_x$, $\mu_y$, or $\mu$) is rather poor, making truncation particularly detrimental to accuracy in this context (especially for low values of the parameter $\epsilon$).

For nonlinear problems, the strategy of bypassing the primal solution is generally not applicable. The linearity of the governing equations and of the QoI functional is conditional of computing the QoI solely from the adjoint solution. In the nonlinear case, the QoI depends nonlinearly on the primal solution, and therefore, one would need either a surrogate model for the primal problem in addition to that for the adjoint, or alternative hyper-reduction techniques to enable fast evaluations.

Finally, the interplay between ROM enrichment and mesh adaptivity is an important topic for ensuring the reliability of simulations. The overall accuracy depends on both the discretization error introduced by the Finite Element basis functions and the reduction error arising from the reduced basis. Being able to assess these two sources of error separately is crucial, as it enables the identification of the dominant error and the design of adaptive strategies, by either refining the mesh or enriching the reduced basis. In this context, adaptive strategies are meant to maintain a balanced control of both errors while ensuring that the surrogate model remains predictive for the QoI~\cite{chamoin2017posteriori}.

\section{Conclusion}
\label{sec:conclusion}

We have introduced an adjoint-based PGD reduced-order model for efficiently and accurately estimating QoIs when dealing with many-query studies. By building the reduced-order model solely for the adjoint problem, repeated primal solves are bypassed. The approach allows for substantial computational savings while maintaining high accuracy. Numerical experiments in the cases of the Poisson problem and of a structural bracket in plane-stress elasticity have demonstrated that the adjoint ROM converges rapidly with respect to the QoI and outperforms the primal ROM in terms of both accuracy and efficiency.
The performance of the proposed approach is explained by the fact that the construction of the surrogate does not take into account the load configurations, enabling fast evaluation across many load cases. The by-product of the method is that it also allows one to quickly create virtual charts to support informed decision-making. Particularly suited to the study of linear problems, the methodology provides a robust, scalable, and efficient tool for many-query analyses. It is especially useful in the early phases of prototyping where quick and reliable QoI predictions are essential. 
Future work will focus on extending the methodology to the case of nonlinear problems and quantities of interest.

\section*{Declarations}

\subsubsection*{Availability of data and materials}
All data generated or analyzed during this study are included in this published article or available from the corresponding author on reasonable request.

\subsubsection*{Competing interests}
The authors declare that they have no known competing financial interests or personal relationships that could have appeared to influence the work reported in this paper.

\subsubsection*{Funding}
Cl\'{e}ment Vella is grateful for the French Tech Emergence Grant awarded by Bpifrance. Serge Prudhomme is grateful for the support by a Discovery Grant from the Natural Sciences and Engineering Research Council of Canada [Grant No.\ RGPIN-2025-07199].

\subsubsection*{Authors' contributions}

\begin{tabularx}{\textwidth}{@{}lX}
\textbf{Cl\'ement Vella}: & Conceptualization, Methodology, Software, Validation, Formal analysis, Investigation, Writing -- original draft, Visualization. \\
\textbf{Serge Prudhomme}: & Conceptualization, Formal analysis, Writing -- review \& editing, Supervision, Funding acquisition, Discussions.
\end{tabularx}

%

\begin{appendices}

\section{}
\label{apx:eval_B}

Let $\Omega = \Omegax \times \Omegay$.
We explicitly evaluate here the bilinear form $B$ using the product of two PGD spatial modes $\varphi = \varphi(x)$, $\forall x \in \Omegax$, and $\psi=\psi(y)$, $\forall y \in \Omegay$, in the case of the Poisson equation.
Let $(\varphi, \psi) \in V_{\mathcal{X}} \times V_{\mathcal{Y}}$ be a pair of spatial modes. Then the bilinear form reads:
\[
B(\varphi \psi, \varphi \psi) = \int_{\Omega}{ \nabla (\varphi \psi) \cdot \nabla (\varphi \psi) \, dx } .
\]
By definition of the gradient:
\[
\nabla (\varphi \psi) = 
\bigg( \dfrac{d\varphi}{dx} \psi, \varphi \dfrac{d\psi}{dy} \bigg),
\]
so that the bilinear form can be expanded as:
\[
\begin{aligned}
B(\varphi \psi, \varphi \psi) &= \int_{\Omegay}{ \int_{\Omegax}{ 
\bigg[ \frac{d\varphi}{dx} \bigg]^{2} \psi^{2} + \varphi^{2} 
\bigg[ \frac{d\psi}{dy} \bigg]^{2} \, dx } dy } \\
&= 
\bigg( \int_{\Omegax}{ \bigg[ \frac{d\varphi}{dx} \bigg]^{2} \, dx } \bigg) 
\bigg( \int_{\Omegay}{ \psi^{2} \, dy } \bigg) + 
\bigg( \int_{\Omegax}{ \varphi^{2} \, dx } \bigg) 
\bigg( \int_{\Omegay}{ \bigg[ \frac{d\psi}{dy} \bigg]^{2} \, dy } \bigg) .
\end{aligned}
\]

\section{}
\label{apx:primal_pgd}

We derive here the PGD approximation of the primal problem in the case of the plane-stress elasticity problem. Given $u_{m-1}$, the mode $(\varphi_m, \lambda_m, \eta_m)$ of the PGD approximation, given in the form: 
\begin{equation}
\label{eq:pgd_ps_elas}
u_{m}(x, \alpha, \beta) = \sum_{i = 1}^{m}{ \varphi_{i}(x) \lambda_{i}(\alpha) \eta_{i}(\beta) },
\end{equation}
satisfies the PGD weak formulation of the parameterized primal problem:
\begin{equation}
\label{frm:param_wr_ps_elas_pri}
\begin{aligned}
& \text{Find}\ (\varphi, \lambda, \eta) \in U \times L^{2}(\mathcal{D}) \times L^{2}(\mathcal{D})\ \text{such that} \\
& \hspace{.1in} \int_{\mathcal{D}^{2}}{ A(\varphi \lambda \eta, v) \, d\alpha \, d\beta} = \int_{\mathcal{D}^{2}}{ L(v) \,  d\alpha \, d\beta} \\
&  \hspace{1.5in} - \int_{\mathcal{D}^{2}}{ A(u_{m - 1}, v) \, d\alpha \, d\beta}, \quad \forall v \in U \times L^{2}(\mathcal{D}) \times L^{2}(\mathcal{D}) ,
\end{aligned}
\end{equation}
with $A(\cdot, \cdot)$ and $L(\cdot)$ as defined in Formulation~\ref{frm:ps_elas}. Assuming that the body force density $f$ is zero everywhere in $\Omega$ and accounting for the Neumann boundary conditions, the linear form $L$ then reads:
\[
L(v) = \int_{\Gammal}{ g_{\alpha} \cdot v \, ds } + \int_{\Gammar}{ g_{\beta} \cdot v \, ds }, \quad \forall v \in U .
\]
Because of the presence of the indicator functions in the bearing loads~\eqref{eq:bearingloads}, $g_{\alpha}$ and $g_{\beta}$ are not separable and need to be approximated by SVDs of their discrete representations.




\end{appendices}


\bibliography{sn-bibliography}

\end{document}